\documentclass[journal=jctcce,manuscript=article]{achemso}

\usepackage[version=3]{mhchem} 
\usepackage{graphicx}
\usepackage{amsmath}
\usepackage{xcolor}
\usepackage[singlelinecheck=false]{caption}
\graphicspath{ {./figs/} }
\usepackage{braket}
\usepackage{array}

\newcolumntype{C}[1]{>{\centering\arraybackslash}p{#1}}

\author{Pavlo Golub}
\author{Andrej Antalik}
\author{Libor Veis}
\author{Jiri Brabec}
\email{jiri.brabec@jh-inst.cas.cz}
\affiliation[jh-inst]
{J. Heyrovsky Institute of Physical Chemistry, v.v.i., Czech Academy of Sciences, Prague, Czech Republic}

\title[An \textsf{achemso} demo]
  {Automatic selection of active spaces for strongly correlated systems 
  using machine learning algorithms}

\begin{document}



\begin{abstract}

The active-space quantum chemical methods could provide very accurate description of strongly correlated electronic systems, which is of tremendous value for natural sciences. The proper choice of the active space is crucial, but a non-trivial task. 
In this article, we present the neural network (NN) based approach for automatic selection of active spaces, focused on transition metal systems.
The training set has been formed from artificial systems composed from one transition metal and various ligands, on which we have performed DMRG and calculated single-site entropy. 

On the selected set of systems, ranging from small benchmark molecules up to larger challenging systems involving two metallic centers, we demonstrate that our ML models could correctly predict the importance of orbitals with the high accuracy.
Also, the ML models show a high degree of transferability on systems much larger than any complex used in training procedures.

\end{abstract}

\section{Introduction}

Material sciences constantly deal with systems involving unusual and intricate electron and magnetic properties. Compounds which include transition metals are among such systems and are widely used in different areas of medicine\cite{Marloye-Berger-2016, Liang-Zhong-at-al-2017, Monro-Colon-2019}, biology\cite{Haas-Franz-2009} and chemistry\cite{Perutz-Procacci-2016}. However, the computational treatment of such strongly correlated systems is inaccurate within purely single-reference wave function approaches due to frequent occurrences of open shells and nearly-degenerated states. Semiempirical methods can be used only very cautiously for these systems\cite{Minenkov-Sharapa-2018} (although with the latest developments structure optimization can be carried out quite accurately\cite{Bursch-Neugebauer-2019}), while density functional theory (DFT) and its extensions together with specially designed hybrid or meta-GGA exchange-correlation functionals allow reasonable accuracy for certain compounds\cite{Buhl-Kabrede-2006,Jensen-Roos-2007,Johnson-Becke-2008,Orio-Pantazis-2009,Tekarli-Drummond-2009,Goodpaster-Barnes-2012,Vlahovich-Peric-2015,Wang-Verma-2018}. Nevertheless, there is no universal approach for selection an appropriate DFT-based scheme and the resulting accuracy can still be behind the corresponding accuracy for main-group elements\cite{Radon-2019,Vogiatzis-Polynski-2019,Chan-Gill-2019}.

\indent The treatment of such problems is often, although not uniquely, seen as the domain of multireference methods, especially those that are based on complete active spaces with initial set of orbitals from a single-reference solution. Active spaces are formed from set of important occupied and virtual orbitals -- i.e. orbitals which capture most of the electron correlation. However, the complexity of the solution generally limits such methods to quite moderate active spaces -- for example, the exact complete active-space self-consistent field (CASSCF) method\cite{Roos-Taylor-1980} suffers from the exponential growth of the active space, that restricts the maximum size approximately up to 20 orbitals. Density matrix renormalization group (DMRG)\cite{White-1992,White-1993,Schollwock-2005} is one of the methods which can operate on larger active spaces and has been used in conjunction with many multireference methods as active space solver to reduce their cost and accurately reproduce both static and dynamic correlations\cite{Zgid-Nooijen-2008,Ghosh-Hachmann-2008,Yanai-Kurashige-2009,Kurashige-Yanai-2011,Nakatani-Guo-2017,Saitow-Kurashige-2013,Zgid-Ghosh-2009,Guo-Watson-2016,Phung-Wouters-2016,Maradzike-Hapka-2020,Faulstich-Mate-2020}. 

\indent In any case, the selection of strongly correlated orbitals and discarding of insignificant ones is an important step in building computational setups for further multireference treatment, since the consideration of all valence space as active space is often beyond the limit of even the most efficient multireference methods. Modern scientific interests are generally in significantly large molecular complexes where the search of an appropriate active spaces should be carried out over hundreds of orbitals. In this case a manual selection is error prone and generally not a good option, especially in automated calculations, for example, when sampling a chemical space. 
That is why a very fast accurate-enough black-box solution is needed.

\indent Recently, several strategies have been proposed for orbital selection. One of them is to go over canonical Hartree-Fock orbitals of the reference solution to natural orbitals (NO)\cite{Lowdin-1955,Davidson-1972}, that are the result of diagonalization of the one-particle density matrix. With this method, active spaces are composed from fractionally occupied NOs, which come either from unrestricted second-order M\o{}ller-Plesset perturbation theory or from the single and doubles configuration interaction\cite{Abrams-Sherrill-2004} approach.
The unrestricted Hartree-Fock reference\cite{Pulay-Hamilton-1988,Keller-Boguslawski-2015} 
experiences problems with excited states and in the case of dominant correlation of a strongly occupied orbital with several other orbitals\cite{Keller-Boguslawski-2015}. The automated scheme based on quasi-NOs from strongly contracted second order \textit{n}-electron valence perturbation theory (SC-NEVPT2\cite{Angeli-Cimiraglia-2001}) has been developed recently\cite{Khedkar-Roemelt-2019,Khedkar-Roemelt-2020}. Quasi-NOs arise when instead of diagonalization of the whole one-particle density matrix only its sub-blocks (internal/internal and external/external parts) are diagonalized. At first the problem is solved self-consistently with CAS or DMRG method on minimal active space, after that the two sub-matrices of the one-particle density matrix are diagonalized and quasi-NOs are used for extension of the active space based on their occupation numbers. The procedure continues iteratively until there are no quasi-NOs with occupational numbers within certain range.

\indent Another strategy\cite{Stein-Reiher-2016,Stein-Reiher-2017,Stein-Reiher-2019} uses cheap, partially converged DMRG calculations with low bond dimension to estimate the inter-orbital entanglement derived from single- and two-orbital von Neumann entropies, which, in turn, are obtained from the eigenvalues of the one- and two-orbital reduced density matrices respectively. Indeed, this measure is independent on previous knowledge of electronic structure of the system and does not significantly dependent on the accuracy of DMRG calculations at least for medium size molecules.\cite{Rissler-Noack-2006,Boguslawski-Tecmer-2012}

\indent The next strategy applies transformations to single-reference molecular orbitals to extract as much useful information as possible. For example, a series of rotations of molecular orbitals in order to maximize their atomic valence character can identify those with significant 3\textit{d} character\cite{Sayfutyarova-Sun-2017}. This might be especially beneficial for coordination complexes.

\indent Naturally, with the rapid increase of the part of machine learning (ML) in material sciences, the strategy appears that utilizes machine learning techniques. Indeed, this approach is appealing since the prediction of active spaces is available almost immediately after the generation of the reference wavefunction, while all other methods involve additional more or less expensive computational steps. In the first study of such kind a set of diatomic molecules with different bond lengths up to the dissociation limit has been subject to ML classification, and the applicability of the approach has been shown\cite{Jeong-Stoneburner-2020}. However, the dependence of ML-based approaches on a proper choice of input data and trainable parameters plus hard-to-ensure transferability of the learned models diminish the possible universality of this strategy. At least for now it can be seen as targeting a specific class of compounds at once. Nevertheless, with a wise choice of ML protocol this class can be sufficiently broad to satisfy a wide range of practical needs.

\indent In this work we devise a neural network (NN) for search of strongly correlated orbitals in transition metal complexes. The test/train set has been chosen to broadly cover metal-ligand interactions. The resulting model can be transferred to larger complexes since it is expected that the metal-ligand interactions and corresponding molecular orbitals contribute the most to the correlation energy. The canonical Hartree-Fock molecular orbitals have been used as input to make the reference as simple as possible. The learning targets (which subsequently are used for construction of active spaces) has been the single-orbital entropies. Finally, the model transferability has been tested with a set of small and medium size molecules which prototypes have not been included to the model at the stage of training (out-of-model molecules).

\section{Theory and method}

\subsection{Density matrix renormalization group}

The density matrix renormalization group (DMRG) method is one of the state-of-the-art approach originally designed for accurate description of linear systems.\cite{White-1992} After the successful application in physics, the approach has been introduced also in theoretical chemistry,\cite{White-Martin-1999,Chan-Head-Gordon-2002, Legeza-Roder-2003} where DMRG is a powerful method for strongly correlated systems requiring large active space, which is impracticable to be treated by other active space methods. 

DMRG can reach full configuration interaction (FCI) limit within the active space, but does not involve dynamical correlation effects from orbitals lying outside of the active space. In order to treat dynamical correlations, in past few years, new methods have been developed, which combine DMRG with other approaches.\cite{Zgid-Nooijen-2008} 

The DMRG algorithm iteratively optimizes the wave function in the matrix product states (MPS) form\cite{SCHOLLWOCK201196}

\begin{equation}
\ket{\Psi_{\rm MPS}} = \sum_{\{n\}} \sum_{\{i\}} A^{n_1}_{i_1}A^{n_2}_{i_1,i_2} ... A^{n_k}_{i_{k-1}} \ket{n_1 n_2...n_k} ,
\label{mps}
\end{equation}

where $k$ is the number of orbitals in the active space. The bounded dimension $M$ of matrices $A$ is so called {\it bond dimension} \cite{White-1992,Schollwock-2005} and determines the accuracy of the DMRG calculation. The renormalized basis optimized during the DMRG process is always truncated, leading to an error, which can be estimated from the so called {\it truncation error} (TRE)\cite{https://doi.org/10.1002/qua.24898}. It can be shown that close to exact energies the DMRG energy can be extrapolated from calculations for continuous series of $M$. 

The accuracy of a prediction of properties based on DMRG (or on any other active space method) strongly depends on the selection of the active space.
The active orbitals are usually selected manually using user's ``intuition", or with respect to the orbital energies, to the entropies, or, by more advantage algorithms.\cite{Stein-Reiher-2016}
The selection based on intuition or on orbital energies, where orbitals are selected as a compact window of highest occupied and lowest unoccupied orbitals, could be very inefficient. Especially when considering important virtual orbitals, which are strongly delocalized and also could lie high above the Fermi level. More sophisticated methods based, for example, on orbital entropy, provide quantum-information based insight, but require an extra computational effort, and involve ``initial" active orbital pre-selection anyway.

The entropy $s^{(N)}$ of the $N$-site system is defined in the sense of the von Neumann entropy as

\begin{equation}
    s^{(N)} = -Tr[\rho^{(N)}ln \rho^{(N)}],
\end{equation}

where $\rho^{(N)}$ is the density matrix for the system.
The single-site entropy $s^{(1)}$, obtained as a trace over all degrees of freedom but one site, quantify how the single orbital is correlated with the rest of the system, i.e. how much quantum information shares with the remainder of set of orbitals. It represents a systematic way how to determine the weight of orbitals inside the active space. If $s^{(1)}$ is small, the orbital is weakly correlated with the others and it indicates that the orbital should not be involved in the active space. Of course, this is true only if the active space already involves some important orbitals, which is usually true if the initial orbitals are pre-selected with respect to orbital energies. 

\subsection{Building machine learning model}

One of the main challenges in building machine learning models is how to choose a proper set of input features. In our case this set should unequivocally describe molecular orbitals from the solution of the relevant multielectron problem. 
We decided to involve descriptors which will allow us to keep the models free of unnecessary dependencies, so
only features, that can be extracted from molecular orbitals alone and multielectron integrals have been considered. One another challenge is to gain an appropriate set of data and find such a way to represent and pre-process it, that will maximally facilitate the ensuing learning procedure. 

\subsubsection{Data generation}

\indent To collect the necessary data we built a set of artificial transition metal complexes by combining seven metals -- Cr, Mn, Fe, Co, Ni, Cu, Zn (the last is not considered as transition metal, however, we decided to include it anyway to generalize the data) -- with ligands (Br\textsuperscript{--}, Cl\textsuperscript{--}, F\textsuperscript{--}, S\textsuperscript{2--}, NH\textsubscript{3}, CO, H\textsubscript{2}O, NC\textsuperscript{--}). The equilibrium metal-ligand distance \textit{x\textsubscript{eq}} has been obtained from rough geometry optimization of molecules at the Hartree-Fock level. Complexes with \textit{x\textsubscript{eq}}, 1.5\textit{x\textsubscript{eq}} and 2\textit{x\textsubscript{eq}} metal-ligand distances have been included to the data set. 
We have considered two, four and six ligands surrounding the metal and it has been assumed that maximum of two different ligands can occur in one complex. We also did not mix different metal-ligand distances, i.e. in single complex all ligands can be only at one type of distance from available three types \textit{x\textsubscript{eq}}, 1.5\textit{x\textsubscript{eq}}, 2\textit{x\textsubscript{eq}}.

\indent Thus totally 5,796 complexes have been obtained. For each complex an active space of 36 orbitals has been chosen, where the number of active electrons has been determined from the sum of valence electrons of constituent metal centres and ligands (Table S1). Assuming that a single orbital is a single sample in machine learning model, this provided in total 208,656 samples - a set large enough to allow a simple train/test/validation split and not to apply a cross-validation procedure. Further for all trainings the data has been randomly split in train and test sets in proportion 3:1, and the train set has been split in actual train set and validation set in proportion 9:1.

\indent The targets have been the corresponding one-orbital entropies, obtained from DMRG calculations with bond dimension $M=1024$ in 16 sweeps. We verified that $s^{(1)}$ obtained for $M=1024$ is very close to $s^{(1)}$ obtained from the same DMRG calculations when using $M=2000$.

\subsubsection{Feature selection and preprocessing}

In this subsection all test trainings have been performed with single hidden layer feedforward NN. Hidden layer has been composed from 1024 neurons with rectified linear unit (ReLU) ($f(z) = max \{ 0,z \} $) as activation function. ReLU also has been chosen as activation function for output layer to ensure the non-negativity of the output values. Dropout regularization procedure\cite{Hinton-Srivastava-at-al-2012, Srivastava-Hinton-at-al-2014} with drop rate 0.1 has been used to reduce overfitting. Trainings have been performed during 2000 epochs and standard, stochastic descent gradient procedure has been used.

\indent We composed the following basic set of input features \textbf{\textit{B}}: single-center one-electron integral, single-center two-electron integral, orbital energy, orbital spatial extension, orbital position label, occupation label, bonding/antibonding label, atomic orbitals encoding. \\

The first four features belong to continuous values space, while the rest are represented by discontinuous values. The bonding/antibonding label is equal to 1 in the case of bonding character of the orbital and -1 in the case of its antibonding character. Occupation label can be 0.25 in the case on virtual orbital, 0.5 in the case of singly occupied orbital and 0.75 in the case of doubly occupied orbital. We decided to prescribe non-zero label to unoccupied orbitals since the absence of electrons might provide an useful information, while allegedly natural label '0' would simply exclude the influence of this feature from the model. Orbital position label describes how far specific molecular orbital from HOMO-LUMO gap. Thus, the first HOMO gets label -1, the second HOMO gets label -2 and so on; the first LUMO gets label 1, the second LUMO gets label 2 and so on. However, to avoid the influence of large numbers (for example, when possible search will be performed over several hundreds of molecular orbitals), the labels have been squeezed to (0,1) range with logistic sigmoid function $\dfrac{1}{1+\textrm{exp}(-k/100)}$, where $k$ is original non-squeezed label.

\indent The last feature describes the types of atomic orbitals that construct specific molecular orbital. It is a one-hot vector with 12 items and has the following mask - [1s 2s 3s 4s 5s 2p 3p 4p 5p 3d 4d 5d]. If some type of atomic orbital is in linear expansion of the molecular orbital (regardless of atom from which it comes), then 1 is inserted in the corresponding place of the one-hot vector, 0 otherwise.

\begin{figure}[h]
\includegraphics[width=14cm]{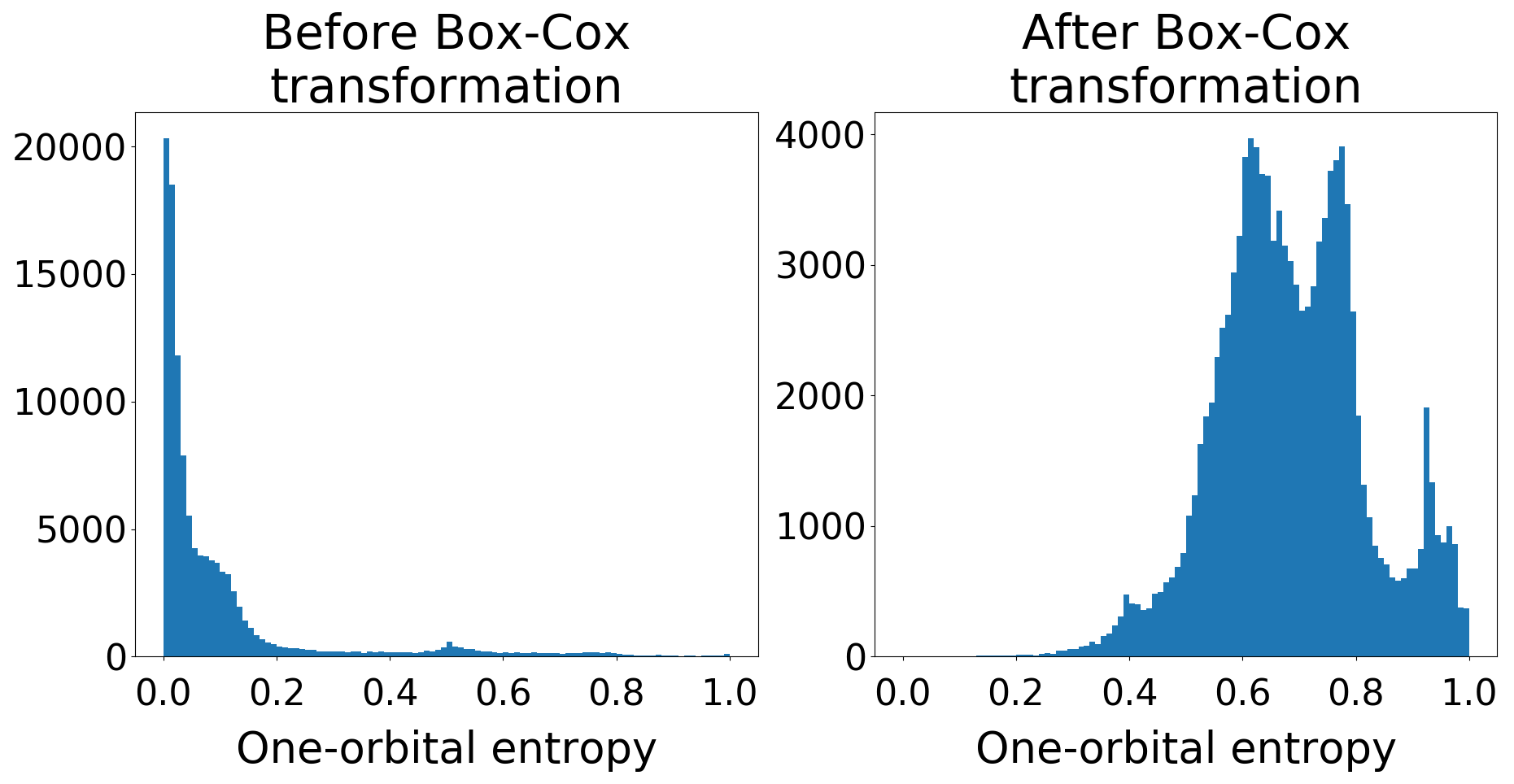}
\caption{Distribution of one-electron entropies before and after the Box-Cox transformation. One-orbital entropies have been squeezed to (0,1) range.}
\label{fig:box-cox-transform}
\centering
\end{figure}

\indent The distribution of target values shows great skewness toward values less than 0.2 (Fig. \ref{fig:box-cox-transform}). Such skewness can affect the model performance, thus the target data was transformed with Box-Cox\cite{Box-Cox-1964} transformation:
\begin{equation} \label{eq:box-cox}
  y(\lambda) = 
    \begin{cases}
      \frac{y^{\lambda}-1}{\lambda} & \quad \lambda \neq 0, \\
      \log y & \quad \lambda = 0.
    \end{cases}
\end{equation}
Indeed, this allowed to improve mean absolute error (MAE) on train data from 0.064 for raw non-transformed data to 0.052 for transformed data, that is almost by 19\%.

\begin{figure}[h]
\includegraphics[width=14cm]{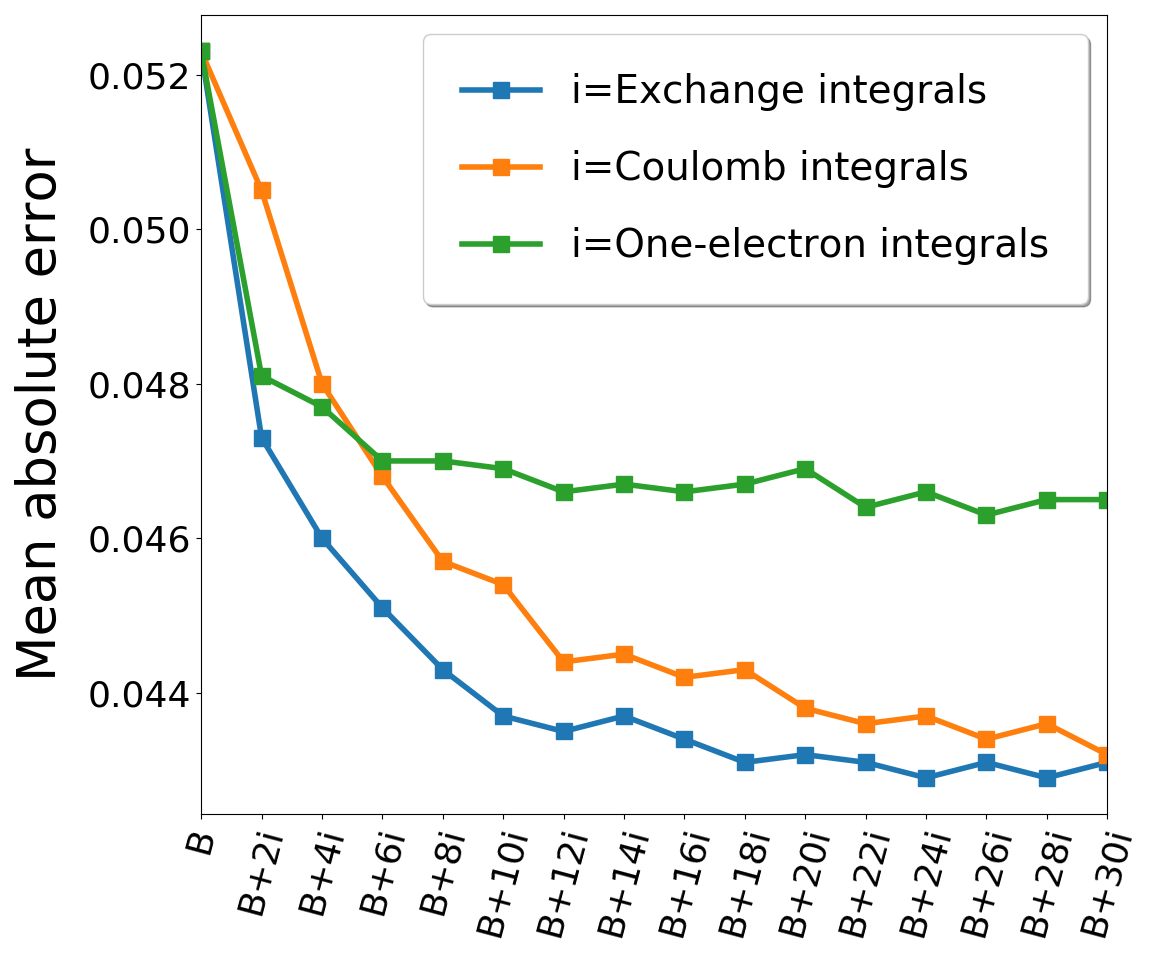}
\caption{The dependence of mean absolute error of train set on the number of largest by absolute value two-centre integrals in input vector. \textbf{\textit{B}} is the basic set of input features, \textit{i} is the corresponding type of integrals. For example, \textit{\textbf{B} + 10i} means that the training was performed on the set of basis features plus 10 largest two-centre integrals of corresponding type in input vector.}
\label{fig:mae_2c_ints1}
\centering
\end{figure}

\indent In recent works 
MO integrals were used in order to optimize orbital ordering, which
allowed to improve the DMRG convergence in certain systems\cite{Chan-Head-Gordon-2002,Legeza-Roder-Hess-2003}, thus it is worth to consider an inclusion of some multicentre one- and two-electron integrals in feature space. However, we considered only two-centre integrals by adding \textit{N} largest of them within respective active space to the input vector. Their values have not been pre-processed, since they range approximately from -2 to 2 without significant outliers. The results of test trainings are at Fig. \ref{fig:mae_2c_ints1}. MAE drops significantly with addition up to 10 two-electron integrals, both Coulomb and exchange, and with addition up to 6 one-electron integrals. However, the addition of more two-centre integrals does not produce so apparent improvement in MAE. Thus the addition of 10-18 Coulomb, 10-18 exchange and 6 one-electron two-centre integrals might be seen as optimal extension of the input feature space in terms of complexity of the model and the gain provided by its inclusion.

\begin{figure}[h]
\includegraphics[width=14cm]{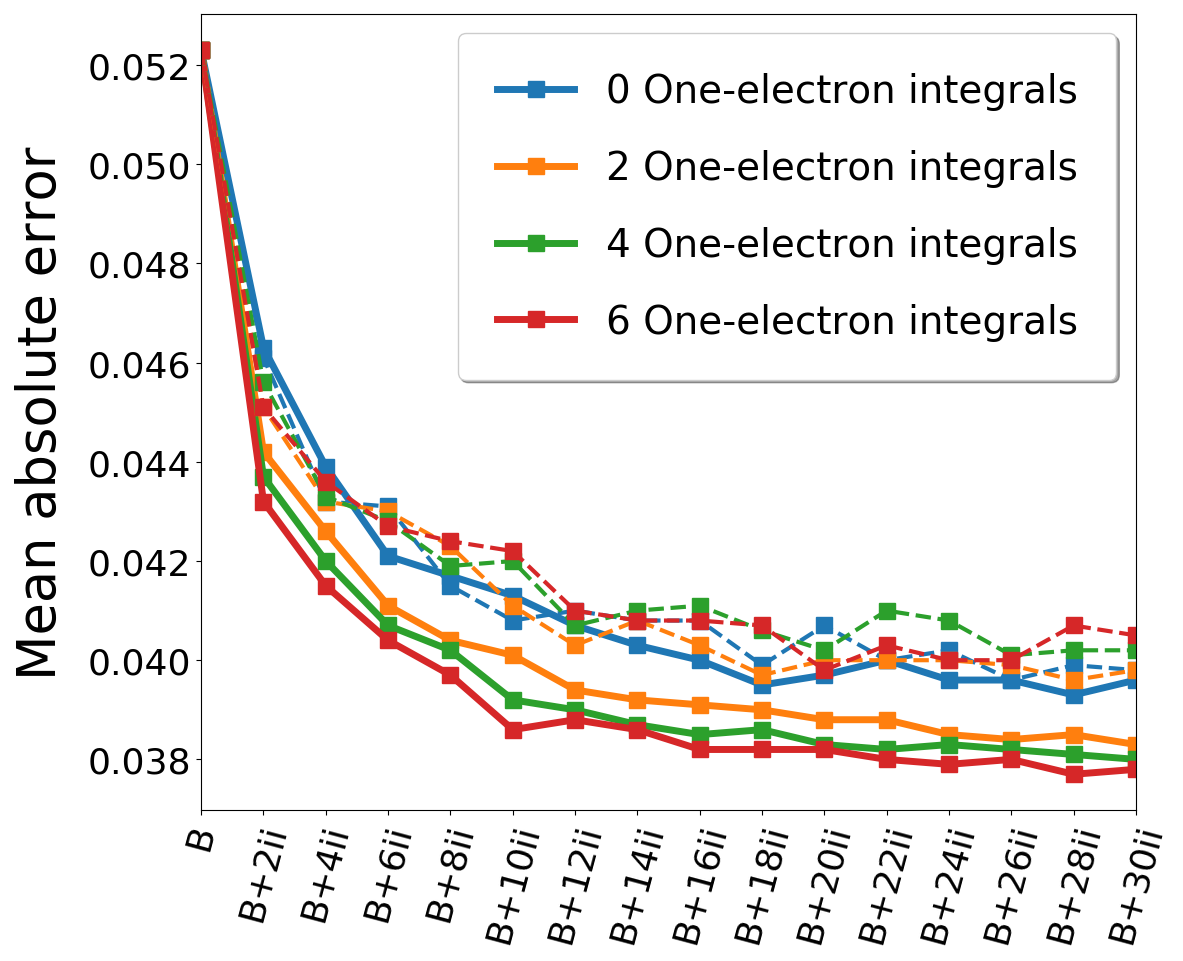}
\caption{The dependence of mean absolute error of train and test sets on the number of largest by absolute value two-centre integrals in input vector. \textbf{\textit{B}} is the basic set of input features, \textit{ii} is the corresponding type of two-electron integrals. For example, \textit{\textbf{B} + 10ii} means that the training was performed on the set of basis features plus 10 largest two-centre Coulomb and exchange integrals plus the number of largest two-centre one-electron integrals as defined by the line color. Solid lines are the train errors, dashed lines are the test errors.}
\label{fig:mae_2c_ints2}
\centering
\end{figure}

\indent However, the the final test with different numbers of one- and two-electron two-centre integrals reveals the appearance of slight overfitting at certain configurations (Fig. \ref{fig:mae_2c_ints2}). The addition to the feature space two-centre one-electron integrals reduces train error, but test error remains approximately the same as for the case without them. In the same time an addition of 18 Coulomb and 18 exchange largest two-centre integrals looks like an optimal extension of the feature space. The drawback might be in the requirement of large enough active spaces to make a proper choice of these integrals. The second drawback is a potential overfitting due to large number of two-electron integrals -- such model can be non-transferable beyond the set of molecules, which have been used for its building. Therefore it has been decided to take only 10 largest two-electron integrals and limit them only to exchange ones. Indeed, in this case MAE drops below 0.044 (figure \ref{fig:mae_2c_ints1}) and improves only by approximately 0.002 with addition of 10 Coulomb integrals (figure \ref{fig:mae_2c_ints2}). In the same time the preference of exchange integrals over Coulomb integrals is obvious from figure \ref{fig:mae_2c_ints1}.

\indent Therefore, the final set of input features has been chosen as the basic features \textbf{\textit{B}} plus 10 largest within the active space two-centre exchange integrals (tcEXC), that is in total 29 features.

\subsubsection{Network architecture}

In the network architecture we decided to keep the ReLU activation function both in hidden and output layers due to its efficiency in computation and sustainability to vanishing gradient problem in comparison to sigmoidal activation functions\cite{Glorot-Bordes-Bengio-2011}. Moreover, ReLU activation function belongs to the family of activations functions, with which an universal approximator based on feedforward network can be build at least in theory\cite{Leshno-Lin-1993}. The dropout regularization procedure with dropout rate 0.1 and stochastic descent gradient procedure from the previous subsection also have been kept. We have performed search over up to 6 hidden layers (thus going from shallow to deep networks) and up to 1536 neurons per layer within the model (number of neurons has been multiply of 16 for facilitate parallel calculations on 16-core processors).

\begin{figure}[h]
\includegraphics[width=14cm]{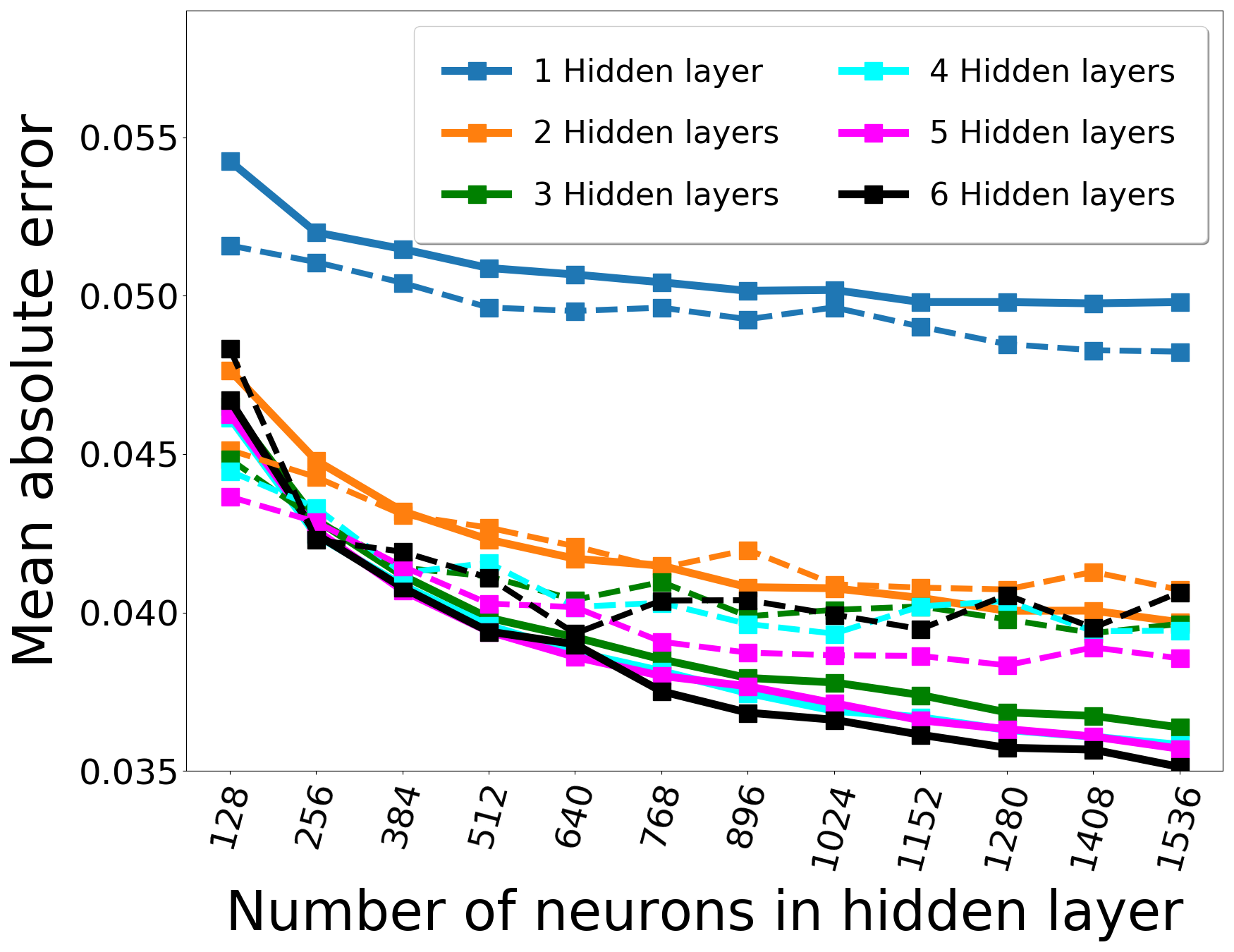}
\caption{The dependence of mean absolute error of train and test sets on the number of hidden layers and neurons within each hidden layer. Solid lines are the train errors, dashed lines are the test errors.}
\label{fig:nn_architecture}
\centering
\end{figure}

\indent MAE on train set seemingly saturates after 3 hidden layers (Fig. \ref{fig:nn_architecture}) and only faintly improves upon transition to a truly deep NNs. In the same time MAE on test data all in all improves up to 5 hidden layers and comes to its best after 896 neurons per layer. The corresponding MAEs for deeper networks are generally larger. 

\begin{figure}[h]
\includegraphics[width=14cm]{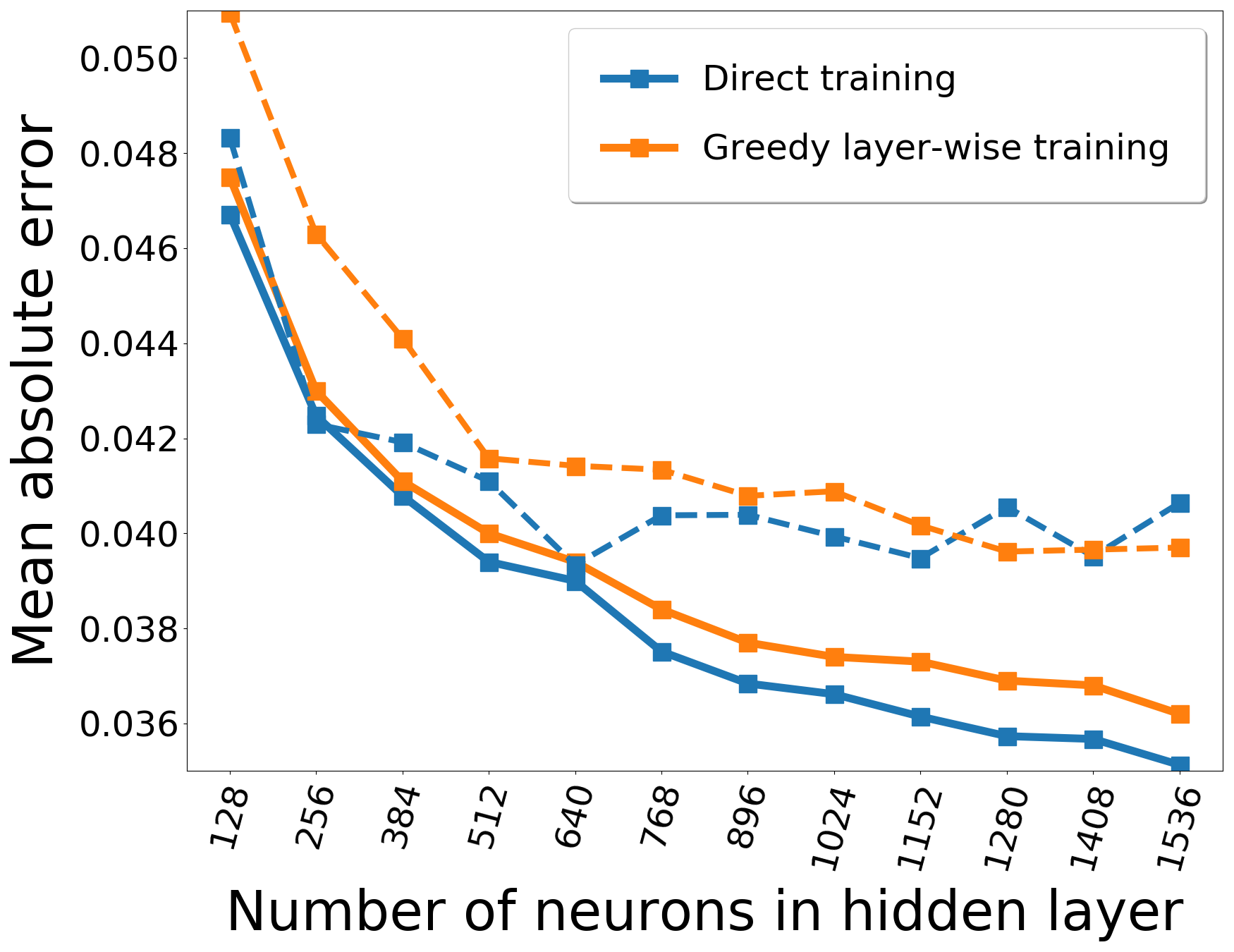}
\caption{The dependence of mean absolute error of train and test sets in 6-hidden layer neural network on the neurons within each hidden layer. Solid lines are the train errors, dashed lines are the test errors.}
\label{fig:nn_6layers}
\centering
\end{figure}

\indent There are a number of alternative ways to improve the performance of deep networks, for example, the greedy layer-wise pre-training technique\cite{Bengio-Lamblin-2006}. With this technique the network is trained by successively adding new hidden layer, the weights and biases of the previous layer are kept frozen thus serving as input parameters to the new layer. In the end trainable parameters over entire network are unfixed and the final fine-tuning training is performed. We have carried out the greedy layer-wise pre-training test on network with 6-hidden layers. The parameters of each new hidden layer have been trained during 1500 epochs with final fine-tuning training during 100 epochs. The results however have been comparable to the case of direct training of the entire network (Fig. \ref{fig:nn_6layers}), thus leaving the conclusions extracted for Fig. \ref{fig:nn_architecture} unchanged.

\indent In the choosing of final model it is desirable to govern not only by the smallest MAE on test set but also by the difference in MAEs on test and train sets. If the last becomes apparent the transferability of the model is likely to worsen. Keeping in mind this considerations we decided to take 5-hidden layered NN with 896 neurons per hidden layer as the reference model. Further it will be referred as \textbf{5x896}. The choice, of course, is not unique and there are other models with comparable performance.

\indent This selected network is still relatively big and could reveal an overfitting problem while testing on out-of-model molecules. Therefore it is useful to evaluate a comparative performance of a lesser NN, which, despite its inferiority in terms of MAE on the model test/train sets, could be more transferable. As example of such smaller NN we have chosen the same network as 5x896 but with only 128 neurons per hidden layer. Further it will be referred as \textbf{5x128}.

\indent It is also worth check if the performance can be altered with an extended input vector. For this purpose the another model with the same architecture but with modified input feature space has been tested. The modification has been in extension of the feature space (\textbf{\textit{B}}+10tcEXC) with one-hot vector encoding atomic types (\textit{AT}). It is a 15-items vector (by the number of all possible atomic species in the test/train complexes: [Cr, Mn, Fe, Co, Ni, Cu, Zn, C, N, O, F, S, Cl, Br, H]) featuring 1 if an atomic orbital from the corresponding atomic specie is involved in construction of the relevant molecular orbital, and 0 otherwise. Further this model will be referred as \textbf{5x896at}. The addition of such vector restricts the applicability of the model to molecules with only specific atoms as constituent items, however this provides a useful comparison to more general 5x896 and 5x128 models.  

\section{Computational details}

All Hartree-Fock self-consistent (SCF) calculations have been performed with NWChem ab initio software package\cite{nwchem-2010} and Ahlrichs pVDZ basis set\cite{ahlrichs-1992}. The respective DMRG calculations have been performed with MOLMPS\cite{brabec2020massively} program, integrated with NWChem multiconfiguration SCF (MCSCF) module to read directly one- and two-electron integrals. NNs have been built and trained with TensorFlow machine learning platform\cite{tensorflow}.

The geometry of ZnBr radical, CuOH, MnCl\textsubscript{2}, Ni(CO)\textsubscript{4}, CrCl\textsubscript{2}O\textsubscript{2}, [Cr\textsubscript{2}O\textsubscript{7}]\textsuperscript{2--}, Fe(C\textsubscript{5}H\textsubscript{5})\textsubscript{2}, [Fe(NO)]\textsuperscript{2+}, trans-[Co(C\textsubscript{2}H\textsubscript{8}N\textsubscript{2})\textsubscript{2}(NO\textsubscript{2})Cl]\textsuperscript{+} molecules have been optimized at DFT level with B3LYP exchange-correlation functional\cite{Stephens-Devlin-1994} and the same Ahlrichs pVDZ basis set. The geometry of oxo-Mn(salen) has been taken from Ref.\cite{Ivanic-Collins-2004} and the singlet geometry of Fe(II)-porphyrin from Ref.\cite{Antalik-Nachtigallova-2020}. The single-orbital entropies have been calculated with bond dimension 2048 until fully convergent with the energy threshold of 10\textsuperscript{-5} a.u. The following active space have been taken (here we use (\textit{a}e, \textit{b}o) notation where \textit{b} is the complete space size and \textit{a} is the number of electrons within this space): ZnBr (17e, 28o), CuOH (18e, 26o), MnCl\textsubscript{2} (21e, 30o), Ni(CO)\textsubscript{4} (42e, 40o), CrCl\textsubscript{2}O\textsubscript{2} (32e, 32o), [Cr\textsubscript{2}O\textsubscript{7}]\textsuperscript{2--} (44e, 44o), Fe(C\textsubscript{5}H\textsubscript{5})\textsubscript{2} (48e, 44o), [Fe(NO)]\textsuperscript{2+} (17e, 30o), oxo-Mn(salen) (52e, 48o), Fe(II)-porphyrin complex (30e, 44o), trans-[Co(C\textsubscript{2}H\textsubscript{8}N\textsubscript{2})\textsubscript{2}(NO\textsubscript{2})Cl]\textsuperscript{+} (70e, 60o), [Fe\textsubscript{2}S\textsubscript{2}(SCH\textsubscript{3})\textsubscript{4}]\textsuperscript{2--} (64e, 54o).

\section{Results and discussion}

\begin{figure}[h]
\includegraphics[width=12cm]{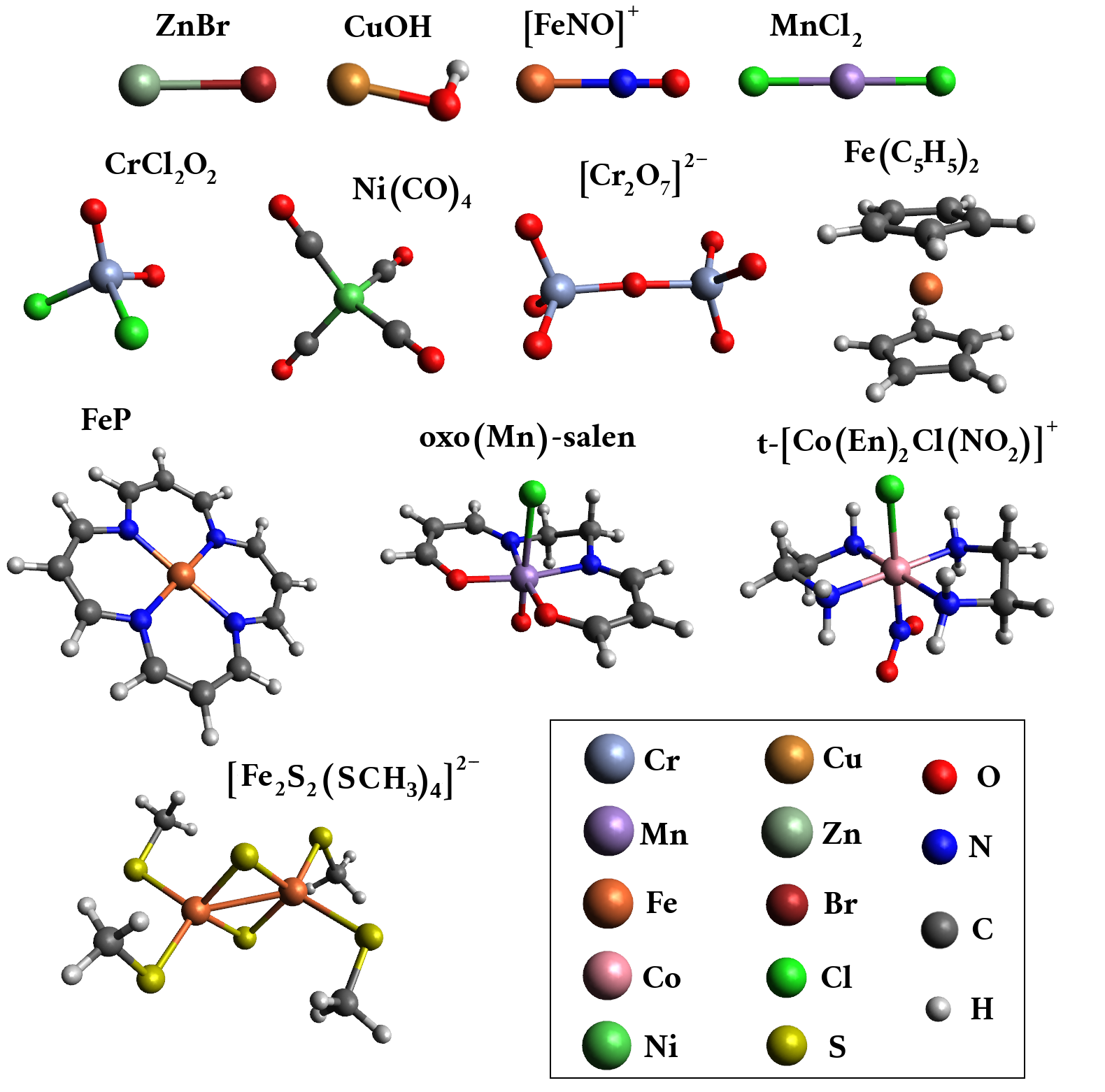}
\caption{Out-of-model molecules.}
\label{fig:molecules}
\centering
\end{figure}

The test set of out-of-model molecules included small and medium size metal complexes: ZnBr radical, CuOH, MnCl\textsubscript{2}, Ni(CO)\textsubscript{4}, CrCl\textsubscript{2}O\textsubscript{2}, [Cr\textsubscript{2}O\textsubscript{7}]\textsuperscript{2--}, Fe(C\textsubscript{5}H\textsubscript{5})\textsubscript{2}, [Fe(NO)]\textsuperscript{2+}, oxo-Mn(salen), trans-[Co(C\textsubscript{2}H\textsubscript{8}N\textsubscript{2})\textsubscript{2}(NO\textsubscript{2})Cl]\textsuperscript{+} (t-Co(En)\textsubscript{2}Cl(NO\textsubscript{2})), singlet Fe(II)-porphyrin model system (FeP), [Fe\textsubscript{2}S\textsubscript{2}(SCH\textsubscript{3})\textsubscript{4}]\textsuperscript{2--} (Fig. \ref{fig:molecules}). Although the prototypes of several molecules (MnCl\textsubscript{2}, Ni(CO)\textsubscript{4}, CrCl\textsubscript{2}O\textsubscript{2}) were among artificial molecules, which have been used to build NN models, in current test set they are optimized (see \textbf{Computational details}) and thus have different geometries.

\indent The accuracy of the models will be evaluated as how many molecular orbitals are correctly predicted to be inside of the active space. The predefined size of the active space is chosen selectively based on the system. From this point of view the corresponding MAE appears as less important evaluation parameter, since, for example, the model may systematically underestimate single-orbital entropies (leading to a relatively large MAE) but in the same time it may correctly predict the content of the active space, which is for us the most important criteria. We will use a shorthand of the type (\textit{a}/\textit{s}) that should be read as such: the model places correctly \textit{a} orbitals to the active space $A$ of size \textit{s}, where $A$ consists of orbitals determined by DMRG single-site entropies. For example, 14/16 means that 14 orbitals were placed correctly to the active space of 16 orbitals, which has been predicted by DMRG. 

We would like to emphasize that even if an $i$-th orbital $a_i$ from $A$ is underestimated by ML model and thus will not fall into $A$, the orbital $b_i$ selected by ML instead could be still relevant if it carries significant contribution from components of $a_i$. 

Also we note that the ordering, which we will refer further, has no relation to the optimal ordering of orbitals in DMRG context, it is just the ordering of orbitals from the one with the largest single-orbital entropy, to the one with the smallest single-orbital entropy. 

Despite of the fact that the {\it orbital importance} is fairly vague term, for the purpose of distinguishing the orbital's weight in the active space we consider orbitals with $s^{(1)} > $ 0.1 as important, for $s^{(1)}$ between 0.05 and 0.1 less important, but still worth of consideration, and below 0.05 as unimportant.

\subsection{Small molecules (ZnBr radical, CuOH, [Fe(NO)]\textsuperscript{2+}).} 

Before we focus on challenging complexes, we have chosen three very simple systems in order to show the performance on different types of metal - ligand interactions.

Zinc has fully occupied 3\textit{d} orbitals and uses 4\textit{s} electrons for bonding. Consequently, in ZnBr radical the most correlated molecular orbitals should be those that are formed from \textit{s} type atomic orbitals of Br and \textit{p} type atomic orbitals of Br and are close to Fermi level. Molecular orbitals which are composed mainly from \textit{d} atomic orbitals should be relatively unimportant. Indeed, there are four molecular orbitals HOMO, HOMO-1, HOMO-2 and HOMO-3 with $s^{(1)}$ larger than 0.1 (Table S2). All three models correctly identify them as the most correlated, moreover, model 5x896 predicts correctly first 7 orbitals, although the orbitals, that go after the first four, can be excluded from the active space without significant influence on accuracy. Also it could be seen that degenerated orbitals are treated correctly and ML models give a consistent prediction among them.

\begin{figure}[ht]
\includegraphics[width=8cm]{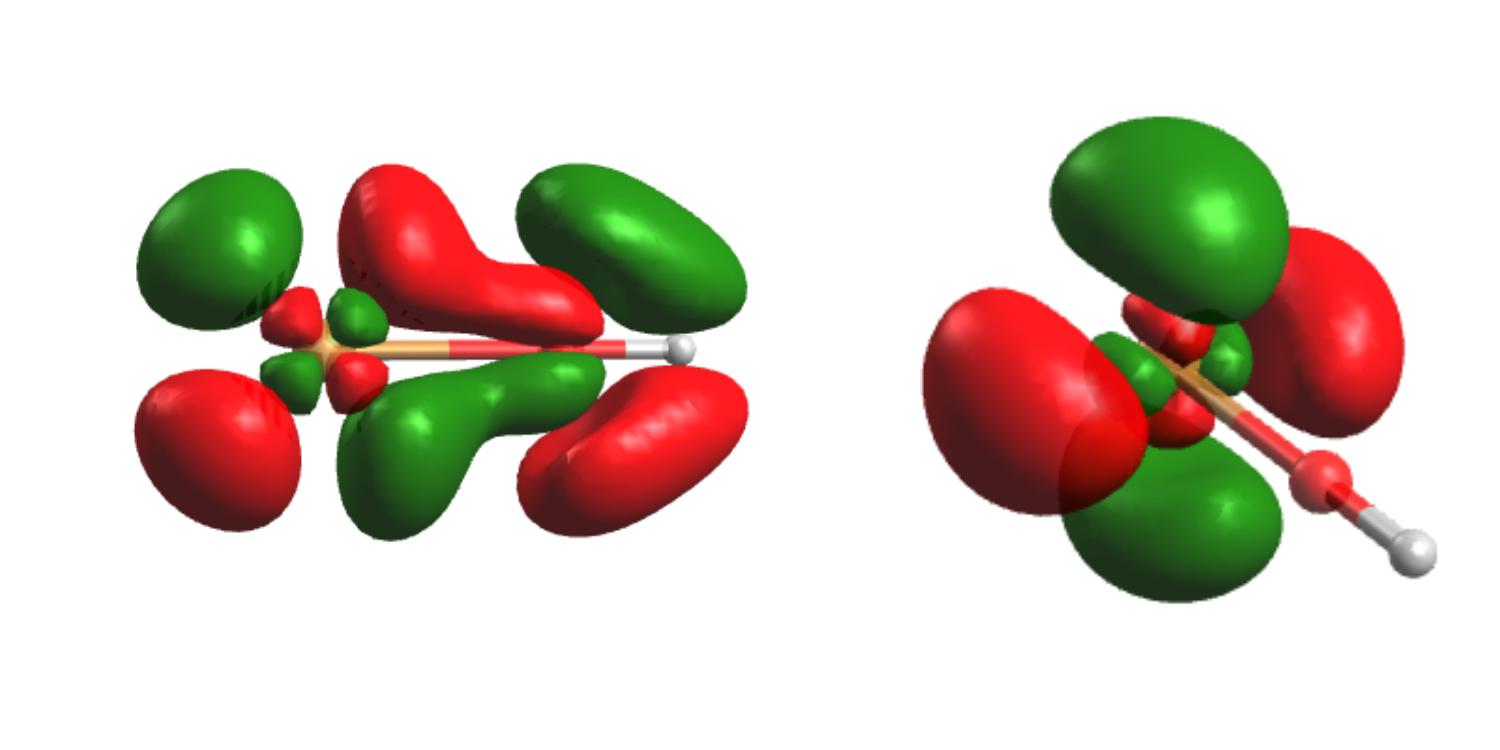}
\caption{An example of $4d$ orbitals (no. 20 and 24 in Table S3) recognised by ML models as important in CuOH system.}
\label{fig:double}
\centering
\end{figure}
\indent A good performance of the models on ZnBr is expected since Zn-Br interaction pattern has been presented in test/train sets abound. A more challenging case is CuOH since hydroxide group (doublet ground-state) has not been used for building of NN models. The transition metal - oxygen interaction has been modelled only with the help of neutral singlet ground-state of the H\textsubscript{2}O ligand. 
The \textbf{basic active space} would be composed from the $3d$ orbitals on Cu and $p-$ orbitals on oxygen, which corresponds to the largest $s^{(1)}$ obtained from DMRG (Table S4). Model 5x896 performs as 5/6, 6/8, 7/10, 9/12, and 12/14 -- the last active space includes all 11 most important (according to DMRG) molecular orbitals that should be enough to account most of the correlation. Model 5x128 misses one orbital from the 11 most important if the active if equal to 14 and finds it only if the active space is extended to 20 orbitals. The same problem is observed for 5x896at model.
On this system, one could nicely see that all three ML models could capture also the \textbf{double shell d} orbitals, which are located higher in the virtual space (especially orbitals 18-21, 23-24, Fig.\ref{fig:double}). These orbitals are slightly underestimated with our models, but still indicating a non-negligible importance. While $s^{(1)}$ by DMRG is between $0.7 - 1.0$, ML models estimate $s^{(1)}$ between $0.57$ and $0.7$.

\indent The [Fe(NO)]\textsuperscript{2+} system represent the biggest challenge from this group. It is not only because metal - nitrosyl was not represented in the set of artificial molecules, but also because the positive charge of the complex assumes the presence a number of unoccupied molecular orbitals below Fermi level -- a rare case in the samples used to build the NN models. As a result models 5x128 and 5x896 tend to underestimate the influence of ligand-based molecular orbitals (i.e. constructed from \textit{s} and \textit{p} atomic orbitals of N and O) and overestimate the influence of virtual molecular orbitals close to Fermi level with admixture of Fe \textit{d} orbitals (Table S4). However, model 5x896 correctly identifies first 6 orbitals and further performs as 9/12, 11/14, and 12/16 (in total there are only 16 orbitals with single-orbital entropies greater or close to 0.1). Even better results are showed by 5x896at model: 10/12, 11/14, and 13/16 since it predicts orbitals 3 -- 5 as more strongly correlated than the models without \textit{AT} encoding does.

\subsection{Model prototypes (MnCl\textsubscript{2}, Ni(CO)\textsubscript{4}, CrCl\textsubscript{2}O\textsubscript{2}).} 

The prototypes of all these molecules have been included to the model test/train sets, but in this test the geometry has been optimized at the DFT/B3LYP level.

From results in Tabs. S5--S7 follow that all strongly correlated orbitals are predicted consistently with only few exceptions -- for example, in CrCl\textsubscript{2}O\textsubscript{2} models 5x896 and 5x896at tend to underestimate orbital number 10 which is a composition of Cl and O \textit{p} orbitals. 

Model 5x896 shows 14/14, 15/16, and 16/18 performances on MnCl\textsubscript{2}; 11/14, 15/16 and 18/18 performances on CrCl\textsubscript{2}O\textsubscript{2}; 10/10, 11/14, 13/16, 15/18, and 17/20 performances on Ni(CO)\textsubscript{4}. The last may seem worse than the previous two, but the model places two orbitals with DMRG single-orbital entropies of 0.118 and 0.141 (numbers 14 and 7 in the active space) instead of the orbitals with DMRG single-orbital entropies of 0.184 and 0.141 (numbers 17 and 8 in the active space), that is apparently an equivalent substitution. Thus there is only one molecular orbital that is seriously misinterpreted. All presented NN models give consistent performance.

\begin{table}[h!]
\centering
\caption{Single-orbital entropies for Fe(C\textsubscript{5}H\textsubscript{5})\textsubscript{2} -- true (DMRG) and predicted by models (5x128, 5x896, 5x896at). The table is cut to show only the most important part of the complete space. The full table can be found in Supporting Materials (Table S8).}
\label{table: ferrocene}
\resizebox{\textwidth}{!}{
\begin{tabular}  {C{0.04\textwidth} C{0.44\textwidth} C{0.1\textwidth} C{0.06\textwidth} C{0.08\textwidth} C{0.08\textwidth} C{0.08\textwidth} C{0.08\textwidth}} 
\hline
\textbf{Orb} & \textbf{Atomic orbitals} & \textbf{Energy} & \textbf{Occ} & \textbf{DMRG} & \textbf{$\textrm{5x128}$} & \textbf{$\textrm{5x896}$} & \textbf{$\textrm{5x896at}$} \\
\hline
3 & \textit{d\textsubscript{xy}}(Fe)    \textit{s}(C)    
 & -0.73927 & 2 & 0.036 & 0.115 & 0.080 & 0.176 \\
\hline
4 & \textit{s}(C)    
 & -0.72808 & 2 & 0.036 & 0.045 & 0.031 & 0.083 \\
\hline
5 & \textit{s}(C)    
 & -0.72804 & 2 & 0.037 & 0.053 & 0.025 & 0.101 \\
\hline
6 & \textit{d\textsubscript{zz}}(Fe)    
 & -0.72408 & 2 & 0.035 & 0.114 & 0.079 & 0.074 \\
\hline
7 & \textit{p\textsubscript{x}}(C)    \textit{p\textsubscript{y}}(C)    
 & -0.70011 & 2 & 0.034 & 0.064 & 0.093 & 0.093 \\
\hline
8 & \textit{d\textsubscript{zz}}(Fe)    \textit{s}(Fe)    \textit{p\textsubscript{z}}(C)    
 & -0.56876 & 2 & 0.091 & 0.150 & 0.141 & 0.116 \\
\hline
9 & \textit{p\textsubscript{y}}(C)    \textit{s}(H)    
 & -0.56082 & 2 & 0.049 & 0.024 & 0.027 & 0.130 \\
\hline
10 & \textit{p\textsubscript{x}}(C)    \textit{s}(H)    
 & -0.56081 & 2 & 0.049 & 0.024 & 0.028 & 0.133 \\
\hline
11 & \textit{p\textsubscript{x}}(C)    \textit{s}(H)    \textit{p\textsubscript{y}}(C)    
 & -0.55409 & 2 & 0.031 & 0.024 & 0.028 & 0.139 \\
\hline
12 & \textit{p\textsubscript{y}}(C)    \textit{s}(H)    \textit{p\textsubscript{x}}(C)    
 & -0.55409 & 2 & 0.031 & 0.024 & 0.028 & 0.136 \\
\hline
13 & \textit{d\textsubscript{xy}}(Fe)    \textit{p\textsubscript{y}}(C)    \textit{p\textsubscript{x}}(C)    
 & -0.52809 & 2 & 0.053 & 0.148 & 0.131 & 0.139 \\
\hline
14 & \textit{p\textsubscript{y}}(C)    \textit{p\textsubscript{x}}(C)    
 & -0.52807 & 2 & 0.052 & 0.095 & 0.190 & 0.121 \\
\hline
15 & \textit{p\textsubscript{y}}(C)    \textit{p\textsubscript{x}}(C)    
 & -0.52585 & 2 & 0.043 & 0.095 & 0.194 & 0.124 \\
\hline
16 & \textit{p\textsubscript{y}}(C)    \textit{p\textsubscript{x}}(C)    
 & -0.52584 & 2 & 0.041 & 0.095 & 0.197 & 0.124 \\
\hline
17 & \textit{d\textsubscript{zz}}(Fe)    \textit{d\textsubscript{xx}}(Fe)    \textit{d\textsubscript{yy}}(Fe)    \textit{s}(Fe)    
 & -0.51161 & 2 & 0.070 & 0.157 & 0.449 & 0.373 \\
\hline
18 & \textit{p\textsubscript{z}}(Fe)    \textit{p\textsubscript{z}}(C)    
 & -0.49987 & 2 & 0.101 & 0.221 & 0.240 & 0.129 \\
\hline
19 & \textit{d\textsubscript{xx}}(Fe)    \textit{d\textsubscript{yy}}(Fe)    \textit{s}(C)    
 & -0.42717 & 2 & 0.267 & 0.118 & 0.203 & 0.132 \\
\hline
20 & \textit{d\textsubscript{xy}}(Fe)    \textit{s}(C)    
 & -0.42716 & 2 & 0.267 & 0.119 & 0.202 & 0.132 \\
\hline
21 & \textit{d\textsubscript{yz}}(Fe)    \textit{p\textsubscript{z}}(C)    
 & -0.34123 & 2 & 0.306 & 0.572 & 0.608 & 0.376 \\
\hline
22 & \textit{d\textsubscript{xz}}(Fe)    \textit{p\textsubscript{z}}(C)    
 & -0.34119 & 2 & 0.308 & 0.580 & 0.618 & 0.381 \\
\hline
23 & \textit{p\textsubscript{z}}(C)    \textit{p\textsubscript{y}}(Fe)    
 & -0.34024 & 2 & 0.159 & 0.497 & 0.210 & 0.241 \\
\hline
24 & \textit{p\textsubscript{z}}(C)    
 & -0.34020 & 2 & 0.161 & 0.538 & 0.219 & 0.422 \\
\hline
25 & \textit{s}(Fe)    \textit{s}(C)    
 & 0.11399 & 0 & 0.014 & 0.020 & 0.015 & 0.027 \\
\hline
26 & \textit{p\textsubscript{z}}(C)    \textit{p\textsubscript{x}}(C)    \textit{s}(H)    
 & 0.20695 & 0 & 0.294 & 0.148 & 0.154 & 0.217 \\
\hline
27 & \textit{p\textsubscript{z}}(C)    \textit{d\textsubscript{xy}}(Fe)    \textit{s}(H)    
 & 0.20697 & 0 & 0.290 & 0.298 & 0.250 & 0.207 \\
\hline
28 & \textit{s}(H)    \textit{s}(C)    
 & 0.21329 & 0 & 0.023 & 0.026 & 0.026 & 0.059 \\
\hline
29 & \textit{s}(H)    \textit{s}(C)    
 & 0.21332 & 0 & 0.023 & 0.026 & 0.026 & 0.059 \\
\hline
30 & \textit{s}(H)    
 & 0.21639 & 0 & 0.022 & 0.038 & 0.034 & 0.051 \\
\hline
31 & \textit{p\textsubscript{z}}(C)    \textit{s}(C)    
 & 0.22002 & 0 & 0.150 & 0.052 & 0.124 & 0.084 \\
\hline
32 & \textit{p\textsubscript{z}}(C)    \textit{s}(C)    
 & 0.22003 & 0 & 0.151 & 0.053 & 0.126 & 0.085 \\
\hline
33 & \textit{s}(H)    \textit{p\textsubscript{x}}(C)    \textit{p\textsubscript{y}}(C)    
 & 0.24266 & 0 & 0.044 & 0.025 & 0.026 & 0.059 \\
\hline
34 & \textit{s}(H)    \textit{p\textsubscript{x}}(C)    
 & 0.24269 & 0 & 0.046 & 0.025 & 0.026 & 0.059 \\
\hline
35 & \textit{d\textsubscript{yz}}(Fe)    \textit{s}(H)    
 & 0.24766 & 0 & 0.203 & 0.601 & 0.173 & 0.171 \\
\hline
36 & \textit{d\textsubscript{xz}}(Fe)    \textit{s}(H)    \textit{p\textsubscript{x}}(C)    
 & 0.24767 & 0 & 0.204 & 0.238 & 0.389 & 0.231 \\
\hline
37 & \textit{s}(C)    \textit{s}(H)    \textit{d\textsubscript{yz}}(Fe)    
 & 0.27184 & 0 & 0.165 & 0.364 & 0.112 & 0.177 \\
\hline
38 & \textit{s}(C)    \textit{s}(H)    
 & 0.27185 & 0 & 0.166 & 0.034 & 0.041 & 0.179 \\
\hline
41 & \textit{s}(Fe)    \textit{s}(H)    
 & 0.27703 & 0 & 0.080 & 0.031 & 0.030 & 0.030 \\
\hline
\end{tabular}}
\end{table}

\begin{figure}[h]
\includegraphics[width=12cm]{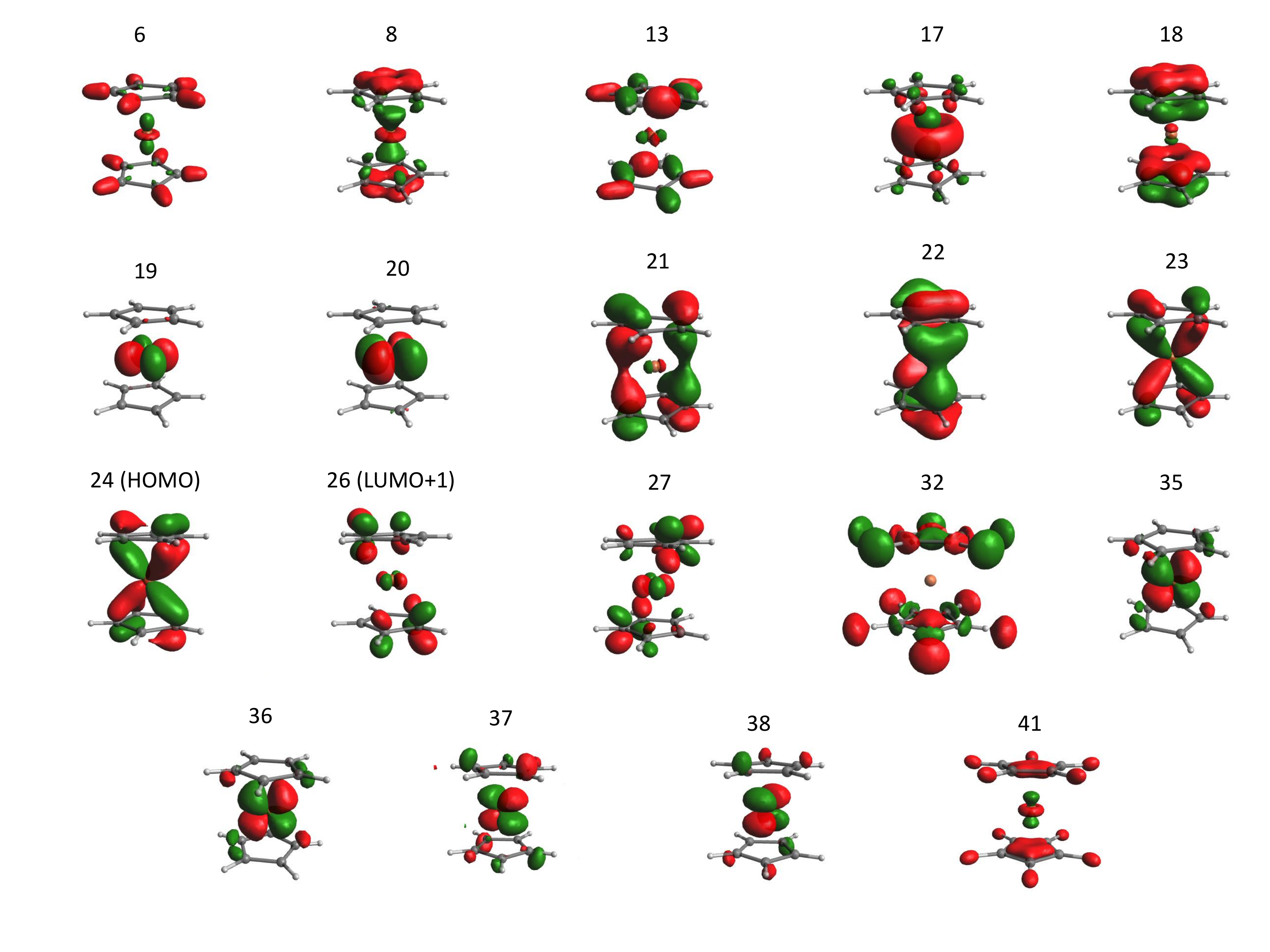}
\caption{The molecular orbitals with the largest one-site entropy for Fe(C\textsubscript{5}H\textsubscript{5})\textsubscript{2}, plus an example of overestimated orbitals 6, 13, 17, and underestimated 41. The numbering corresponds to the orbital number in table \ref{table: ferrocene}.}
\label{fig:ferrocene}
\centering
\end{figure}

\subsection{Challenging systems}

\textbf{Ferrocene (Fe(C\textsubscript{5}H\textsubscript{5})\textsubscript{2}).} 
Despite its seeming simplicity, ferrocene represents a system with a complicated electronic structure. All C atoms from the two pentadienyl rings (Cp) interact equally with the metal center, donating 10 $\pi$ electrons to it to complete the outer shell of Kr. In the same time ferrocene preserves aromatic character of the penta-ring and can be involved in the same reactions as benzene\cite{Bhatt-2016}. There are two main conformers of ferrocene -- eclipse and staggered; the eclipse conformation is considered as the global minimum, while staggered conformation is deemed as a saddle point\cite{Bohn-Haaland-1966, Xu-Xie-2003, Mohammadi-Ganesan-2012}. Nevertheless, the energy difference between them is small and in present investigation the staggered ferrocene has been chosen. The main route to its computational characterization is an adequate account for correlation effects, and in the same time the molecular structure of ferrocene can be successfully predicted on the basis of single-reference wave function\cite{Park-Almlof-1991, Koch-Jorgensen-1996}. The molecular orbitals in the active space should account both for bonding and antibonding interaction between ligands and metal center, and for \textit{d} -- \textit{d} correlation on the metal\cite{Pierloot-Persson-1995}. The active space traditionally is filled with molecular orbitals which are constructed from bonding and antibonding combinations of Cp $\pi$ and Fe 3\textit{d} orbitals (\textit{d\textsubscript{xz}}, \textit{d\textsubscript{yz}}) to account for Fe-Cp covalent interactions, the orbitals which are in charge for backdonation from formally double occupied 3\textit{d} (\textit{d\textsubscript{xy}}, \textit{d\textsubscript{x\textsuperscript{2}--y\textsuperscript{2}}}) orbitals into Cp $\pi$* orbitals, and the rest orbitals with dominant 3\textit{d} character (\textit{d\textsubscript{z\textsuperscript{2}}} and correlated to it orbitals) \cite{Pierloot-Persson-1995, Vancoillie-Zhao-2011}.

Selected orbitals are visualized in Fig.\ref{fig:ferrocene}.
All MO orbitals with combined \textit{d}/\textit{p}(Fe) and \textit{p}(C) AO contributions are recognized as important. Moreover, models expand the active space with the orbitals constructed mostly from \textit{p}(C) AOs (like molecular orbitals number 26, 27, 31, 32), which seemingly might be skipped by manual selection. Model 5x896 performs as 11/16, 14/18 and 16/20, however, it includes to the active space (size 16) another orbitals like 17, which also has a strong component from Fe \textit{d}. It should be noted that model 5x896at, in contrast to 5x128 and 5x896, overestimates the row of orbitals from 9 to 16.

\begin{table}[h!]
\centering
\caption{Single-orbital entropies for [Cr\textsubscript{2}O\textsubscript{7}]\textsuperscript{2--} -- true (DMRG) and predicted by models (5x128, 5x896, 5x896at). The table is cut to show only the most important part of the complete space. Label O{\color{blue}\textsuperscript{B}} marks bridging oxygen atom. The full table can be found in Supporting Materials (Table S9).}
\label{table: dichromate}
\resizebox{\textwidth}{!}{
\begin{tabular}  {C{0.04\textwidth} C{0.44\textwidth} C{0.1\textwidth} C{0.06\textwidth} C{0.08\textwidth} C{0.08\textwidth} C{0.08\textwidth} C{0.08\textwidth}} 
\hline
\textbf{Orb} & \textbf{Atomic orbitals} & \textbf{Energy} & \textbf{Occ} & \textbf{DMRG} & \textbf{$\textrm{5x128}$} & \textbf{$\textrm{5x896}$} & \textbf{$\textrm{5x896at}$} \\
\hline
1 & \textit{s}(O)    \textit{p\textsubscript{x}}(Cr)    
 & -0.91158 & 2 & 0.025 & 0.024 & 0.029 & 0.016 \\
\hline
2 & \textit{d\textsubscript{xy}}(Cr)    \textit{p\textsubscript{y}}(O{\color{blue}\textsuperscript{B}})    \textit{p\textsubscript{x}}(O{\color{blue}\textsuperscript{B}})    \textit{d\textsubscript{yy}}(Cr)    \textit{d\textsubscript{zz}}(Cr)    
 & -0.32459 & 2 & 0.166 & 0.162 & 0.113 & 0.352 \\
\hline
3 & \textit{d\textsubscript{yz}}(Cr)    \textit{d\textsubscript{xz}}(Cr)    \textit{p\textsubscript{z}}(O)    
 & -0.31079 & 2 & 0.214 & 0.176 & 0.548 & 0.278 \\
\hline
4 & \textit{d\textsubscript{xx}}(Cr)    \textit{d\textsubscript{yz}}(Cr)    \textit{d\textsubscript{zz}}(Cr)    \textit{p\textsubscript{x}}(O)    
 & -0.31074 & 2 & 0.212 & 0.220 & 0.505 & 0.425 \\
\hline
5 & \textit{d\textsubscript{yz}}(Cr)    \textit{d\textsubscript{xz}}(Cr)    \textit{p\textsubscript{z}}(O)    
 & -0.30683 & 2 & 0.206 & 0.188 & 0.565 & 0.304 \\
\hline
6 & \textit{d\textsubscript{xx}}(Cr)    \textit{d\textsubscript{zz}}(Cr)    \textit{d\textsubscript{yz}}(Cr)    \textit{d\textsubscript{xy}}(Cr)    \textit{p\textsubscript{x}}(O)    \textit{p\textsubscript{y}}(O)    
 & -0.30683 & 2 & 0.205 & 0.254 & 0.707 & 0.449 \\
\hline
7 & \textit{d\textsubscript{xy}}(Cr)    \textit{d\textsubscript{xz}}(Cr)    \textit{d\textsubscript{yy}}(Cr)    \textit{p\textsubscript{z}}(O{\color{blue}\textsuperscript{B}})    \textit{p\textsubscript{x}}(O{\color{blue}\textsuperscript{B}})    
 & -0.28589 & 2 & 0.178 & 0.198 & 0.726 & 0.324 \\
\hline
8 & \textit{d\textsubscript{xz}}(Cr)    \textit{p\textsubscript{z}}(O{\color{blue}\textsuperscript{B}})    \textit{d\textsubscript{xy}}(Cr) \textit{p\textsubscript{x}}(O)   \textit{p\textsubscript{x}}(O{\color{blue}\textsuperscript{B}})    \textit{d\textsubscript{yz}}(Cr)    
 & -0.28581 & 2 & 0.178 & 0.198 & 0.727 & 0.324 \\
\hline
9 & \textit{d\textsubscript{yy}}(Cr)    \textit{d\textsubscript{yz}}(Cr)    \textit{p\textsubscript{z}}(O)    \textit{d\textsubscript{xx}}(Cr)    \textit{d\textsubscript{xz}}(Cr)    \textit{d\textsubscript{xy}}(Cr)    
 & -0.25806 & 2 & 0.223 & 0.299 & 0.963 & 0.443 \\
\hline
10 & \textit{d\textsubscript{yz}}(Cr)    \textit{d\textsubscript{xz}}(Cr)    \textit{p\textsubscript{x}}(O)    \textit{d\textsubscript{xy}}(Cr)    \textit{p\textsubscript{z}}(O)    
 & -0.25804 & 2 & 0.222 & 0.264 & 0.708 & 0.386 \\
\hline
11 & \textit{d\textsubscript{xy}}(Cr)    \textit{p\textsubscript{y}}(O)    \textit{p\textsubscript{x}}(O)    \textit{s}(Cr)    
 & -0.25749 & 2 & 0.189 & 0.233 & 0.641 & 0.189 \\
\hline
12 & \textit{s}(Cr)    \textit{p\textsubscript{y}}(O) \textit{p\textsubscript{y}}(O{\color{blue}\textsuperscript{B}})   \textit{p\textsubscript{x}}(O)    
 & -0.22898 & 2 & 0.124 & 0.577 & 0.307 & 0.070 \\
\hline
13 & \textit{p\textsubscript{z}}(O) \textit{p\textsubscript{z}}(O{\color{blue}\textsuperscript{B}})   \textit{p\textsubscript{y}}(O)    
 & -0.19940 & 2 & 0.108 & 0.523 & 0.213 & 0.122 \\
\hline
14 & \textit{p\textsubscript{x}}(O)  \textit{p\textsubscript{x}}(O{\color{blue}\textsuperscript{B}})  \textit{p\textsubscript{z}}(O)    
 & -0.19938 & 2 & 0.110 & 0.524 & 0.213 & 0.122 \\
\hline
15 & \textit{s}(Cr)    \textit{p\textsubscript{z}}(O)    \textit{p\textsubscript{y}}(O)    
 & -0.17430 & 2 & 0.187 & 0.329 & 0.254 & 0.081 \\
\hline
16 & \textit{s}(Cr)    \textit{p\textsubscript{z}}(O)    \textit{p\textsubscript{y}}(O)    \textit{p\textsubscript{x}}(O)    
 & -0.17271 & 2 & 0.186 & 0.339 & 0.246 & 0.080 \\
\hline
17 & \textit{p\textsubscript{y}}(O)    \textit{p\textsubscript{x}}(O)    
 & -0.16543 & 2 & 0.156 & 0.192 & 0.201 & 0.140 \\
\hline
18 & \textit{p\textsubscript{y}}(O)    \textit{p\textsubscript{x}}(O)    
 & -0.16539 & 2 & 0.154 & 0.191 & 0.201 & 0.132 \\
\hline
19 & \textit{p\textsubscript{z}}(O)    \textit{p\textsubscript{x}}(O)    \textit{p\textsubscript{y}}(O)    
 & -0.14091 & 2 & 0.118 & 0.195 & 0.228 & 0.110 \\
\hline
20 & \textit{p\textsubscript{z}}(O)    \textit{p\textsubscript{x}}(O)    \textit{p\textsubscript{y}}(O)    
 & -0.14084 & 2 & 0.120 & 0.214 & 0.222 & 0.111 \\
\hline
21 & \textit{p\textsubscript{z}}(O{\color{blue}\textsuperscript{B}})    \textit{p\textsubscript{y}}(O)    
 & -0.12653 & 2 & 0.148 & 0.217 & 0.209 & 0.088 \\
\hline
22 & \textit{p\textsubscript{x}}(O) \textit{p\textsubscript{x}}(O{\color{blue}\textsuperscript{B}})   \textit{p\textsubscript{y}}(O)    
 & -0.12646 & 2 & 0.153 & 0.221 & 0.208 & 0.089 \\
\hline
23 & \textit{s}(Cr)    \textit{d\textsubscript{zz}}(Cr)    \textit{d\textsubscript{xx}}(Cr)    \textit{d\textsubscript{yy}}(Cr)    
 & 0.31417 & 0 & 0.006 & 0.011 & 0.013 & 0.010 \\
\hline
24 & \textit{s}(Cr)    \textit{d\textsubscript{yy}}(Cr)    \textit{d\textsubscript{xx}}(Cr)    \textit{d\textsubscript{zz}}(Cr)    
 & 0.32091 & 0 & 0.005 & 0.009 & 0.008 & 0.005 \\
\hline
25 & \textit{d\textsubscript{xz}}(Cr)    \textit{d\textsubscript{yz}}(Cr)    \textit{d\textsubscript{yy}}(Cr)    \textit{d\textsubscript{zz}}(Cr)    
 & 0.33435 & 0 & 0.309 & 0.595 & 0.462 & 0.372 \\
\hline
26 & \textit{d\textsubscript{xy}}(Cr)    \textit{d\textsubscript{yy}}(Cr)    \textit{d\textsubscript{xx}}(Cr)    \textit{d\textsubscript{yz}}(Cr)    \textit{d\textsubscript{xz}}(Cr)    \textit{p\textsubscript{y}}(O)    
 & 0.33436 & 0 & 0.313 & 0.174 & 0.403 & 0.155 \\
\hline
27 & \textit{d\textsubscript{xy}}(Cr)    \textit{s}(O{\color{blue}\textsuperscript{B}}))    \textit{d\textsubscript{yy}}(Cr)    \textit{d\textsubscript{zz}}(Cr)    \textit{d\textsubscript{yz}}(Cr)    
 & 0.34395 & 0 & 0.260 & 0.599 & 0.518 & 0.198 \\
\hline
28 & \textit{d\textsubscript{xy}}(Cr)    \textit{d\textsubscript{xz}}(Cr)    \textit{d\textsubscript{yy}}(Cr)    \textit{d\textsubscript{xx}}(Cr)    \textit{d\textsubscript{yz}}(Cr)    
 & 0.37184 & 0 & 0.286 & 0.325 & 0.359 & 0.211 \\
\hline
29 & \textit{d\textsubscript{xz}}(Cr)    \textit{d\textsubscript{yz}}(Cr)    \textit{d\textsubscript{xy}}(Cr)    \textit{d\textsubscript{zz}}(Cr)    \textit{d\textsubscript{yy}}(Cr)    
 & 0.37193 & 0 & 0.284 & 0.464 & 0.553 & 0.304 \\
\hline
30 & \textit{d\textsubscript{zz}}(Cr)    \textit{d\textsubscript{xy}}(Cr)    \textit{d\textsubscript{xx}}(Cr)    \textit{p\textsubscript{x}}(O)    \textit{d\textsubscript{yz}}(Cr)    
 & 0.40265 & 0 & 0.300 & 0.297 & 0.765 & 0.198 \\
\hline
31 & \textit{d\textsubscript{xz}}(Cr)    \textit{d\textsubscript{yz}}(Cr)    \textit{p\textsubscript{z}}(O)    
 & 0.40270 & 0 & 0.300 & 0.327 & 0.275 & 0.184 \\
\hline
32 & \textit{d\textsubscript{xx}}(Cr)    \textit{d\textsubscript{yz}}(Cr)    \textit{d\textsubscript{zz}}(Cr)    \textit{p\textsubscript{x}}(O)    \textit{p\textsubscript{z}}(O)    \textit{d\textsubscript{xy}}(Cr)    
 & 0.41318 & 0 & 0.312 & 0.169 & 0.343 & 0.131 \\
\hline
33 & \textit{d\textsubscript{yz}}(Cr)    \textit{d\textsubscript{xz}}(Cr)    \textit{p\textsubscript{z}}(O)    
 & 0.41329 & 0 & 0.314 & 0.309 & 0.250 & 0.184 \\
\hline
34 & \textit{d\textsubscript{xy}}(Cr)    \textit{p\textsubscript{y}}(O{\color{blue}\textsuperscript{B}})    \textit{d\textsubscript{yy}}(Cr)    \textit{d\textsubscript{zz}}(Cr)    
 & 0.44228 & 0 & 0.237 & 0.397 & 0.958 & 0.173 \\
 \hline
35 & \textit{s}(Cr)    \textit{s}(O)    \textit{d\textsubscript{yy}}(Cr)    \textit{d\textsubscript{xx}}(Cr)    
 & 0.57797 & 0 & 0.016 & 0.018 & 0.016 & 0.020 \\
\hline
\end{tabular}}
\end{table}

\begin{figure}[ht]
\includegraphics[width=12cm]{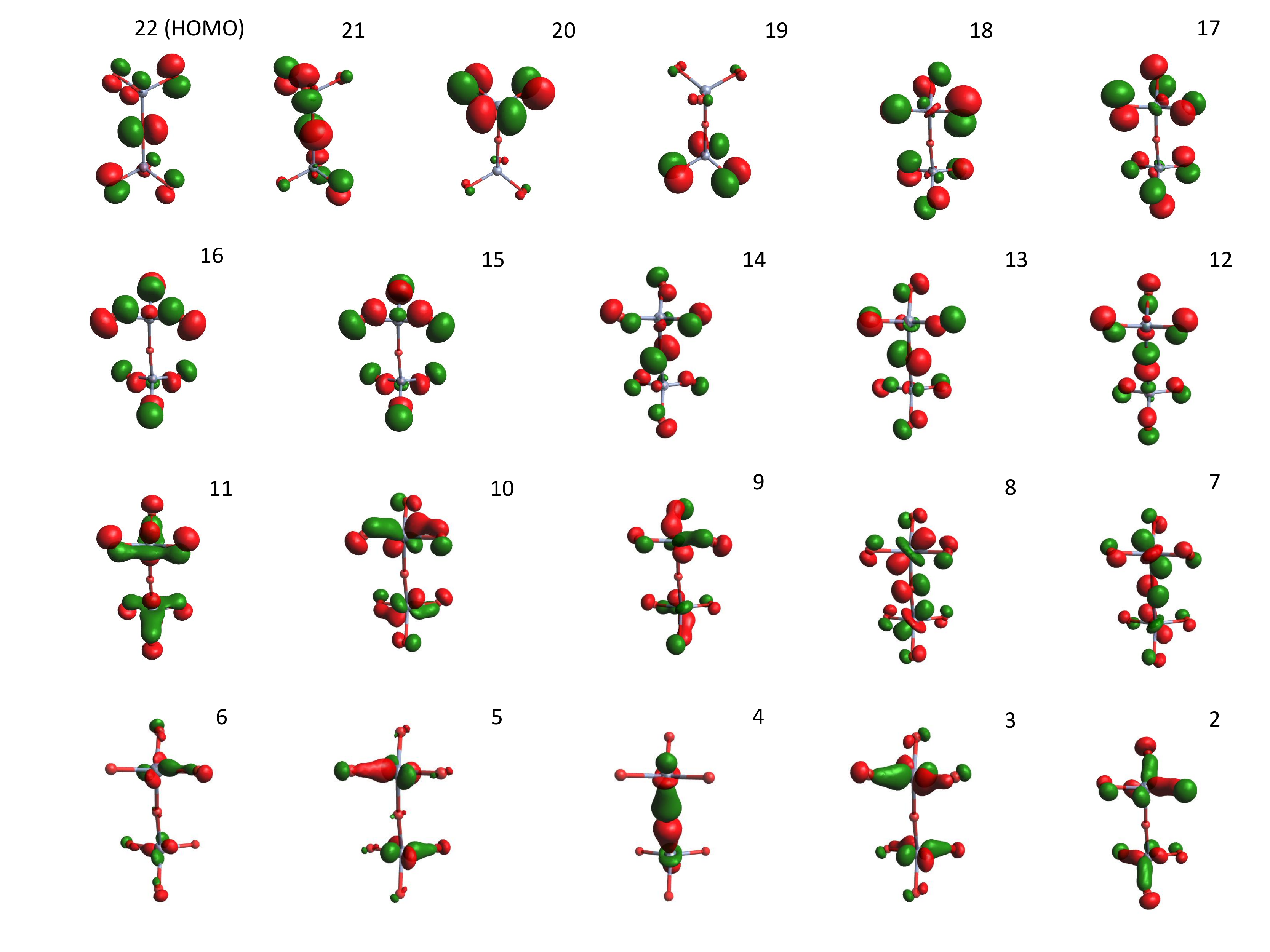}
\caption{The occupied molecular orbitals for [Cr\textsubscript{2}O\textsubscript{7}]\textsuperscript{2--}. The numbering is in accordance with the numbering of orbitals in table \ref{table: dichromate}.}
\label{fig:dichromate}
\centering
\end{figure}

\begin{figure}[ht]
\includegraphics[width=12cm]{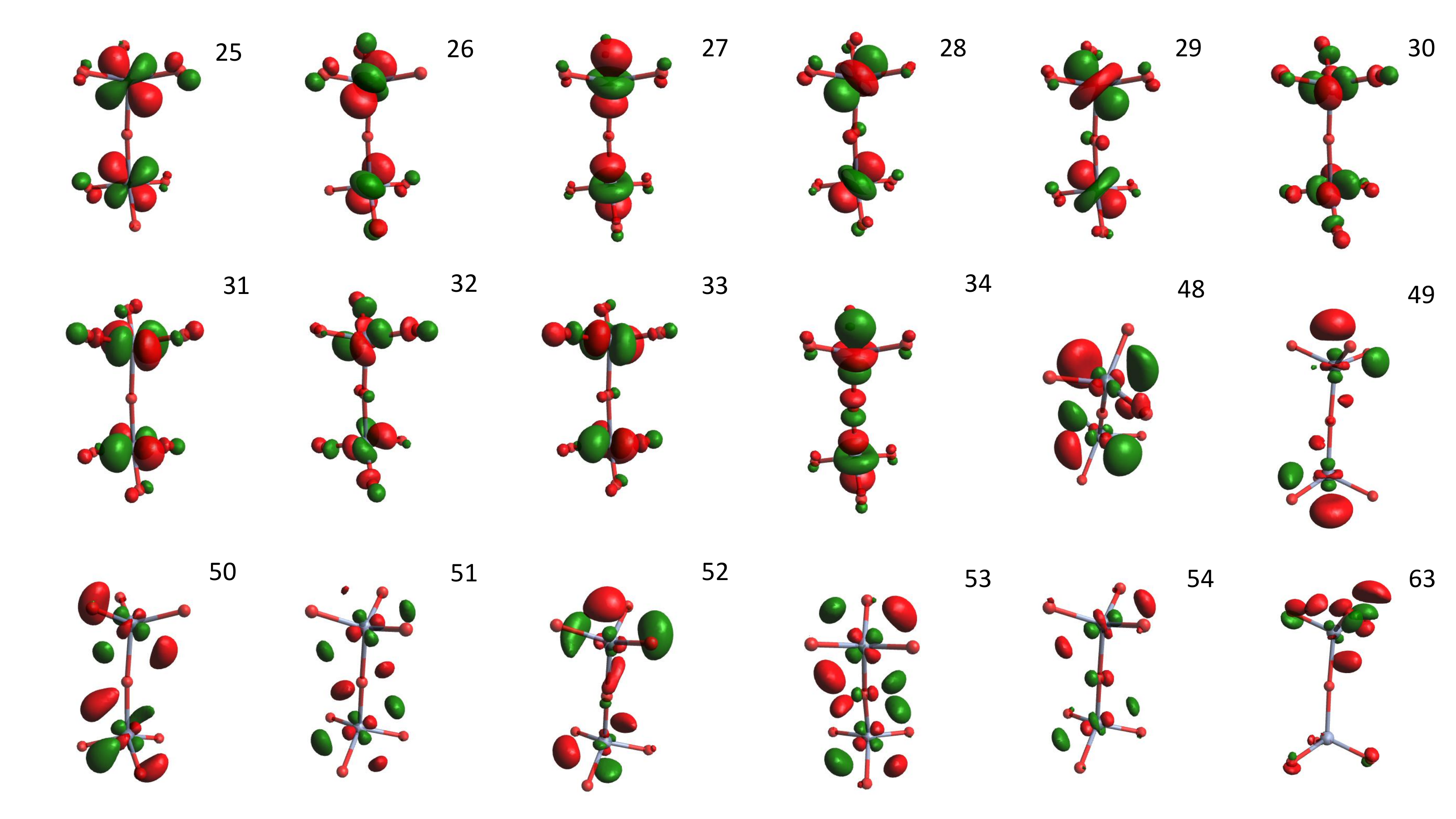}
\caption{The virtual molecular orbitals for [Cr\textsubscript{2}O\textsubscript{7}]\textsuperscript{2--}. The numbering is in accordance with the numbering of orbitals in table \ref{table: dichromate}.}
\label{fig:dichromate2}
\centering
\end{figure}

\textbf{[Cr\textsubscript{2}O\textsubscript{7}]\textsuperscript{2--}.} Dichromate dianion has two Cr(VI) atoms connected by bridging oxygen and characterizes by a ligand-to-metal charge transfer transition\cite{Mesters-Duran-1993}. The picture is complicated not only with several different types of Cr-O (Cr$=$O, Cr--O\textsuperscript{-}) bonds but also with Cr-Cr indirect cooperation, not covered by artificial samples in NN models. The strongly delocalized MO orbitals are vizualized on Figs. \ref{fig:dichromate} and \ref{fig:dichromate2}. The task to chose an active space of size 20 is greatly challenging since there are 31 orbitals with DMRG single-orbital entropies larger than 0.1 (Table \ref{table: dichromate}) and clearly a larger active space is needed to fully account for correlation. The models 5x896 and 5x128 absolutely correctly prescribe to all of them large singe-orbital entropies. Model 5x896at is again inferior in this sense, since it missed 4 orbitals. However, in comparison to other insignificant orbitals those 4 misinterpreted orbitals have larger predicted single-orbital entropies. Thus, all models show performance of 31/31.  It may be emphasized, that the models do not blindly prescribe large entropies to molecular orbitals with a significant \textit{d} contribution from metals. Insignificant orbitals (like 23, 24 and 35, 36, 38 (Table S9)) are also regarded by the models as such. Moreover, the selection over a much larger space leads to the same set of orbitals in the active space (see Table S10 for the space of 100 orbitals). The models do not overestimate the importance of high lying virtual orbitals and deep lying occupied orbitals. There is only one case with the model 5x896at, when it interprets low lying occupied orbital as more correlated than the aforementioned 4 orbitals. There might be difficulties with the selection of a small active space, since, for example, model 5x896 places the most correlated orbital (number 33) at the 21st position. However, as has been mentioned above, small active spaces are not an adequate option for such strongly correlated system.
It is worth of note that ML models 5x128 and 5x896 indicate the importance of some double-d shell orbitals no. 60-61, 63-64 (Tab. S10).

\begin{table}[h!]
\centering
\caption{Single-orbital entropies for Fe(II)-porphyrin model system -- true (DMRG) and predicted by models (5x128, 5x896, 5x896at). The table is cut to show only the most important part of the complete space. The full table can be found in Supporting Materials (Table S11).}
\label{table: porf}
\resizebox{\textwidth}{!}{
\begin{tabular}  {C{0.04\textwidth} C{0.44\textwidth} C{0.1\textwidth} C{0.06\textwidth} C{0.08\textwidth} C{0.08\textwidth} C{0.08\textwidth} C{0.08\textwidth}} 
\hline
\textbf{Orb} & \textbf{Atomic orbitals} & \textbf{Energy} & \textbf{Occ} & \textbf{DMRG} & \textbf{$\textrm{5x128}$} & \textbf{$\textrm{5x896}$} & \textbf{$\textrm{5x896at}$} \\
\hline
3 & \textit{p\textsubscript{x}}(C)    \textit{p\textsubscript{y}}(C)    
 & -0.520 & 2 & 0.008 & 0.052 & 0.065 & 0.065 \\
\hline
4 & \textit{p\textsubscript{z}}(N)    
 & -0.473 & 2 & 0.146 & 0.169 & 0.173 & 0.140 \\
\hline
5 & \textit{p\textsubscript{x}}(N)    \textit{p\textsubscript{y}}(N)    
 & -0.464 & 2 & 0.027 & 0.134 & 0.147 & 0.097 \\
\hline
6 & \textit{p\textsubscript{y}}(N)    \textit{p\textsubscript{x}}(N)    
 & -0.464 & 2 & 0.027 & 0.135 & 0.148 & 0.097 \\
\hline
7 & \textit{d\textsubscript{xy}}(Fe)    \textit{p\textsubscript{x}}(N)    \textit{p\textsubscript{y}}(N)    
 & -0.460 & 2 & 0.175 & 0.190 & 0.200 & 0.135 \\
\hline
8 & \textit{p\textsubscript{z}}(C)    
 & -0.418 & 2 & 0.153 & 0.192 & 0.194 & 0.235 \\
\hline
9 & \textit{d\textsubscript{xx}}(Fe)    \textit{d\textsubscript{yy}}(Fe)    
 & -0.368 & 2 & 0.101 & 0.287 & 0.401 & 0.180 \\
\hline
10 & \textit{p\textsubscript{z}}(C)    \textit{d\textsubscript{xz}}(Fe)    
 & -0.343 & 2 & 0.243 & 0.128 & 0.754 & 0.725 \\
\hline
11 & \textit{p\textsubscript{z}}(C)    \textit{d\textsubscript{yz}}(Fe)    
 & -0.343 & 2 & 0.255 & 0.127 & 0.744 & 0.726 \\
\hline
12 & \textit{d\textsubscript{zz}}(Fe)    \textit{d\textsubscript{xx}}(Fe)    \textit{d\textsubscript{yy}}(Fe)    \textit{s}(Fe)    
 & -0.284 & 2 & 0.246 & 0.619 & 0.824 & 0.989 \\
\hline
13 & \textit{p\textsubscript{z}}(C)    \textit{p\textsubscript{z}}(N)    
 & -0.268 & 2 & 0.283 & 0.153 & 0.920 & 0.431 \\
\hline
14 & \textit{d\textsubscript{yz}}(Fe)    \textit{p\textsubscript{z}}(C)    
 & -0.148 & 2 & 0.765 & 0.308 & 0.828 & 0.534 \\
\hline
15 & \textit{d\textsubscript{xz}}(Fe)    \textit{p\textsubscript{z}}(C)    
 & -0.148 & 2 & 0.770 & 0.311 & 0.829 & 0.537 \\
\hline
16 & \textit{p\textsubscript{z}}(C)    
 & -0.015 & 0 & 0.422 & 0.104 & 0.089 & 0.169 \\
\hline
17 & \textit{d\textsubscript{yz}}(Fe)    \textit{p\textsubscript{z}}(C)    
 & 0.116 & 0 & 0.832 & 0.194 & 0.252 & 0.208 \\
\hline
18 & \textit{d\textsubscript{xz}}(Fe)    \textit{p\textsubscript{z}}(C)    
 & 0.116 & 0 & 0.828 & 0.195 & 0.252 & 0.207 \\
\hline
19 & \textit{s}(Fe)    \textit{s}(C)    
 & 0.128 & 0 & 0.035 & 0.022 & 0.013 & 0.011 \\
\hline
20 & \textit{p\textsubscript{z}}(N)    \textit{p\textsubscript{z}}(C)    
 & 0.137 & 0 & 0.251 & 0.138 & 0.147 & 0.149 \\
\hline
21 & \textit{s}(H)    
 & 0.177 & 0 & 0.003 & 0.021 & 0.021 & 0.021 \\
\hline
22 & \textit{s}(H)    \textit{s}(C)    
 & 0.192 & 0 & 0.003 & 0.015 & 0.013 & 0.024 \\
\hline
23 & \textit{s}(H)    \textit{s}(C)    
 & 0.192 & 0 & 0.004 & 0.015 & 0.013 & 0.024 \\
\hline
24 & \textit{s}(H)    \textit{s}(C)    
 & 0.213 & 0 & 0.016 & 0.014 & 0.010 & 0.026 \\
\hline
25 & \textit{s}(H)    \textit{p\textsubscript{y}}(C)    \textit{p\textsubscript{x}}(C)    \textit{s}(C)    
 & 0.223 & 0 & 0.001 & 0.019 & 0.017 & 0.022 \\
\hline
26 & \textit{p\textsubscript{z}}(Fe)    \textit{p\textsubscript{z}}(C)    
 & 0.236 & 0 & 0.121 & 0.135 & 0.122 & 0.054 \\
\hline
27 & \textit{s}(H)    \textit{s}(C)    
 & 0.243 & 0 & 0.002 & 0.015 & 0.015 & 0.024 \\
\hline
28 & \textit{s}(H)    \textit{s}(C)    
 & 0.243 & 0 & 0.002 & 0.015 & 0.015 & 0.024 \\
\hline
29 & \textit{p\textsubscript{z}}(C)    
 & 0.253 & 0 & 0.163 & 0.159 & 0.190 & 0.058 \\
\hline
30 & \textit{s}(H)    \textit{p\textsubscript{y}}(C)    \textit{p\textsubscript{x}}(C)    \textit{s}(C)    
 & 0.258 & 0 & 0.001 & 0.020 & 0.018 & 0.022 \\
\hline
31 & \textit{p\textsubscript{z}}(C)    \textit{p\textsubscript{z}}(N)    \textit{d\textsubscript{yz}}(Fe)    
 & 0.277 & 0 & 0.209 & 0.154 & 0.190 & 0.142 \\
\hline
32 & \textit{p\textsubscript{z}}(C)    \textit{p\textsubscript{z}}(N)    \textit{d\textsubscript{xz}}(Fe)    
 & 0.277 & 0 & 0.183 & 0.155 & 0.192 & 0.143 \\
\hline
... &   
 &   &   &   &   &   &   \\
\hline
41 & \textit{p\textsubscript{z}}(Fe)    \textit{p\textsubscript{z}}(N)    \textit{p\textsubscript{z}}(C)    
 & 0.386 & 0 & 0.039 & 0.058 & 0.073 & 0.060 \\
\hline
42 & \textit{d\textsubscript{xy}}(Fe)    \textit{s}(C)    
 & 0.462 & 0 & 0.212 & 0.141 & 0.380 & 0.068 \\
\hline
\end{tabular}}
\end{table}

\begin{figure}[ht]
\includegraphics[width=12cm]{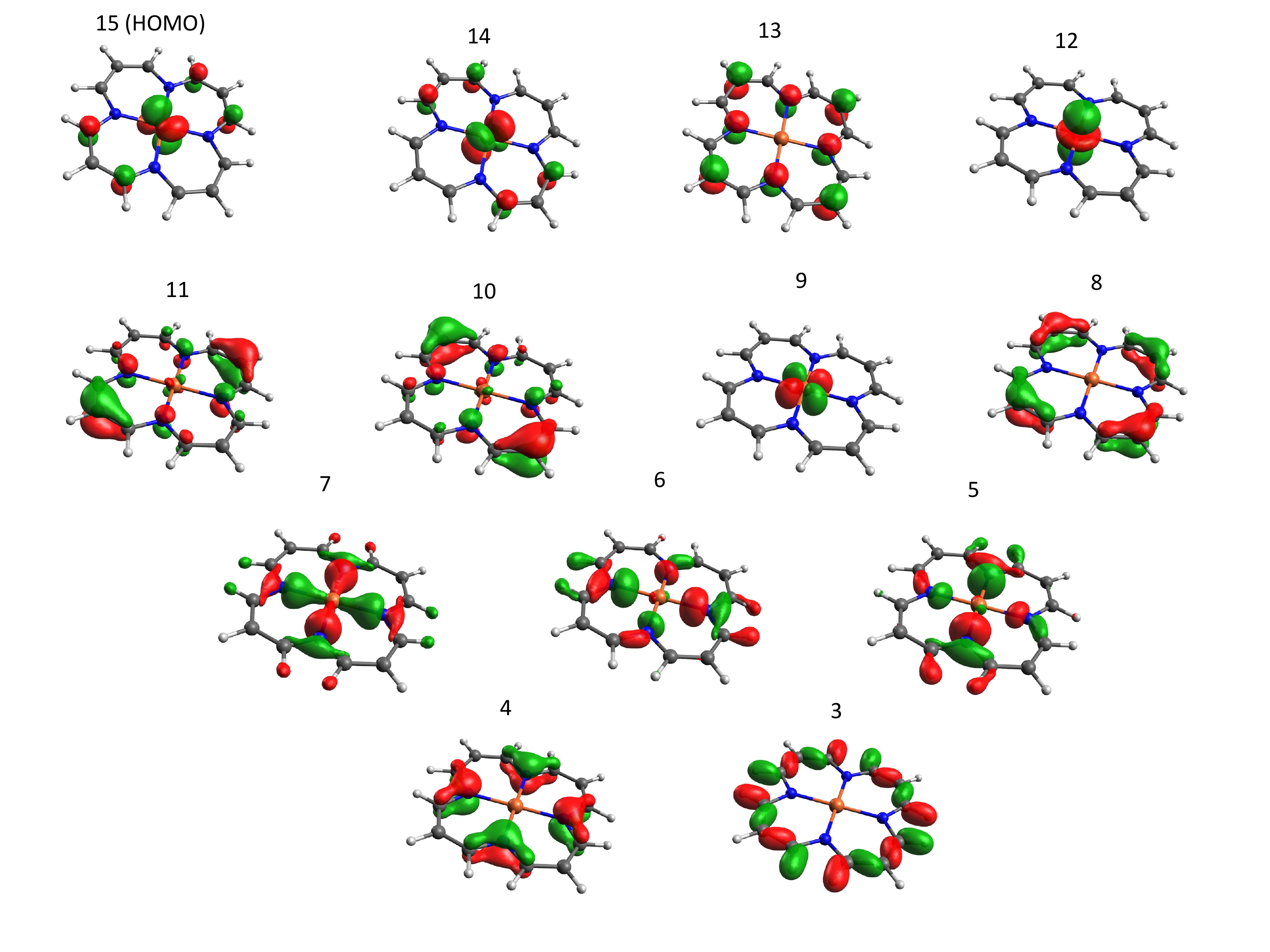}
\caption{The occupied space for Fe(II)-porphyrin model system. The numbering is in accordance with the numbering of orbitals in table \ref{table: porf}.}
\label{fig:porf}
\centering
\end{figure}
\begin{figure}[ht]
\includegraphics[width=12cm]{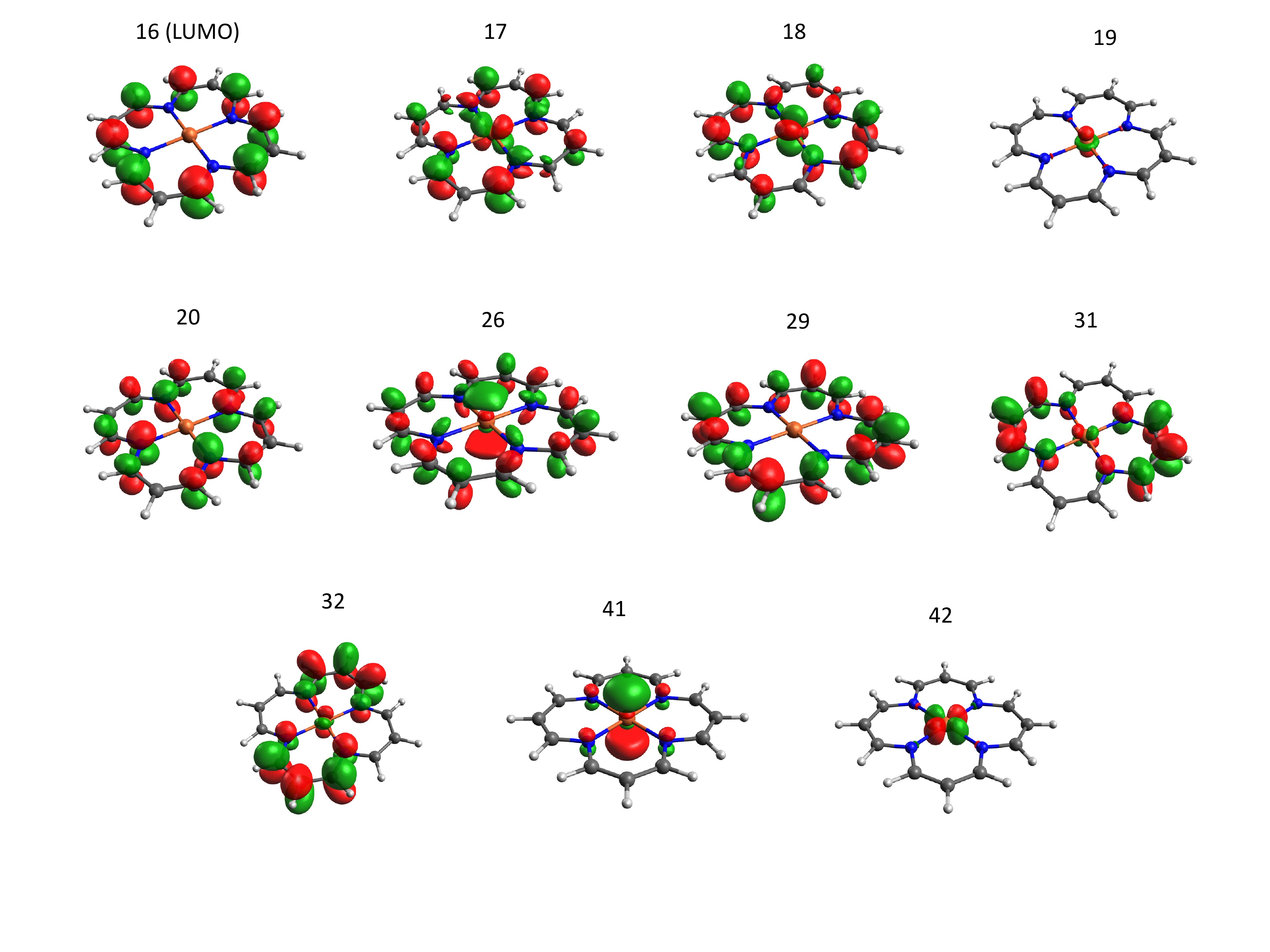}
\caption{The virtual space for Fe(II)-porphyrin model system. The numbering is in accordance with the numbering of orbitals in table \ref{table: porf}.}
\label{fig:porf2}
\centering
\end{figure}

\textbf{Fe(II)-porphyrin model system.} 

The recent studies of the porphyrin complexes have been performed with large active spaces, including all Fe 3\textit{d} and 4\textit{d} orbitals, and a set of $\sigma$ and $\pi$ orbitals from porphyrin ring that leads to larger active spaces of sizes 38 or 44\cite{Olivares-Amaya-Hu-2015, Smith-Mussard-2017, Manni-Alavi-2018, Manni-Kats-2019, Levine-Hait-2020}. However, this is a necessary option, since smaller active spaces fail to correctly predict the most energetically favourable spin configuration\cite{Smith-Mussard-2017, Manni-Kats-2019, Levine-Hait-2020}.

In our calculations we counted 19 orbitals with large single-entropy values $s^{(1)} > 0.1$. They range from orbitals near HOMO/LUMO with significant \textit{d} and \textit{p} characters from Fe and C correspondingly, to low- and highly lying orbitals with \textit{p} characters from N and \textit{d} from Fe (Figs.\ref{fig:porf} and \ref{fig:porf2}). Apparently, they are responsible for Fe-N and indirect Fe-C interactions. Only one of them is missed by model 5x896 when selecting the active space of 20 orbitals (in general it performs as 18/20, missing also the 20th by importance orbital), but all of them are in when the active space of 22 orbitals is chosen (20/22). This can be considered as a very good performance, since other orbitals have much smaller single-orbital entropies and should not contribute much to the correlation. Similarly perform other two models -- 5x128 as 18/20, 5x896at as 17/20. The most challenging orbital is 16 (Fig.\ref{fig:porf2}, it has mainly $\pi$ character of C and is HOMO in Hartree-Fock calculation, Fig. \ref{fig:porf}), which is apparently accounts for conjugacy of the system -- pattern, not included to the NN model. However, as has been mentioned, in accordance to all tested models it still gets in the active space of size 22.

\begin{table}[h!]
\centering
\caption{Single-orbital entropies for oxo(Mn)-salen -- true (DMRG) and predicted by models (5x128, 5x896, 5x896at). The table is cut to show only the most important part of the complete space. The full table can be found in Supporting Materials (Table S12).}
\label{table: oxosalen}
\resizebox{\textwidth}{!}{
\begin{tabular}  {C{0.04\textwidth} C{0.44\textwidth} C{0.1\textwidth} C{0.06\textwidth} C{0.08\textwidth} C{0.08\textwidth} C{0.08\textwidth} C{0.08\textwidth}} 
\hline
\textbf{Orb} & \textbf{Atomic orbitals} & \textbf{Energy} & \textbf{Occ} & \textbf{DMRG} & \textbf{$\textrm{5x128}$} & \textbf{$\textrm{5x896}$} & \textbf{$\textrm{5x896at}$} \\
\hline
1 & \textit{p\textsubscript{x}}(C)    \textit{s}(H)    \textit{p\textsubscript{z}}(C)    
 & -0.703 & 2 & 0.076 & 0.021 & 0.022 & 0.050 \\
\hline
2 & \textit{p\textsubscript{y}}(O)    \textit{d\textsubscript{xz}}(Mn)    \textit{p\textsubscript{y}}(C)    \textit{p\textsubscript{x}}(O)    
 & -0.698 & 2 & 0.057 & 0.091 & 0.064 & 0.107 \\
\hline
3 & \textit{p\textsubscript{y}}(O)    \textit{d\textsubscript{yz}}(Mn)    \textit{p\textsubscript{y}}(N)    
 & -0.681 & 2 & 0.214 & 0.099 & 0.100 & 0.091 \\
\hline
4 & \textit{p\textsubscript{y}}(O)    \textit{d\textsubscript{xy}}(Mn)    
 & -0.676 & 2 & 0.364 & 0.117 & 0.110 & 0.145 \\
\hline
5 & \textit{d\textsubscript{xz}}(Mn)    \textit{d\textsubscript{yy}}(Mn)    \textit{d\textsubscript{xy}}(Mn)    \textit{d\textsubscript{yz}}(Mn)    \textit{d\textsubscript{zz}}(Mn)    \textit{p\textsubscript{z}}(O)    \textit{p\textsubscript{x}}(O)    \textit{d\textsubscript{xx}}(Mn)    
 & -0.669 & 2 & 0.111 & 0.133 & 0.139 & 0.210 \\
\hline
6 & \textit{p\textsubscript{y}}(O)    \textit{p\textsubscript{y}}(C)    \textit{d\textsubscript{yz}}(Mn)    
 & -0.666 & 2 & 0.128 & 0.118 & 0.107 & 0.133 \\
\hline
7 & \textit{d\textsubscript{yy}}(Mn)    \textit{d\textsubscript{xy}}(Mn)    \textit{d\textsubscript{xz}}(Mn)    \textit{p\textsubscript{z}}(C)    \textit{d\textsubscript{zz}}(Mn)    \textit{p\textsubscript{x}}(C)    \textit{s}(H)    
 & -0.645 & 2 & 0.045 & 0.019 & 0.021 & 0.083 \\
\hline
8 & \textit{d\textsubscript{zz}}(Mn)    \textit{d\textsubscript{xx}}(Mn)    \textit{p\textsubscript{z}}(O)    \textit{d\textsubscript{xz}}(Mn)    
 & -0.638 & 2 & 0.466 & 0.171 & 0.271 & 0.343 \\
\hline
9 & \textit{d\textsubscript{zz}}(Mn)    \textit{d\textsubscript{xx}}(Mn)    \textit{p\textsubscript{z}}(O)    \textit{p\textsubscript{z}}(C)    \textit{s}(H)    
 & -0.620 & 2 & 0.243 & 0.042 & 0.026 & 0.059 \\
\hline
10 & \textit{p\textsubscript{y}}(C)    \textit{s}(H)    
 & -0.606 & 2 & 0.083 & 0.017 & 0.019 & 0.116 \\
\hline
11 & \textit{s}(H)    \textit{p\textsubscript{x}}(C)    \textit{p\textsubscript{z}}(N)    \textit{p\textsubscript{z}}(C)    
 & -0.596 & 2 & 0.093 & 0.020 & 0.023 & 0.095 \\
\hline
12 & \textit{s}(H)    \textit{p\textsubscript{y}}(C)    \textit{p\textsubscript{y}}(O)    
 & -0.567 & 2 & 0.046 & 0.017 & 0.021 & 0.187 \\
\hline
13 & \textit{p\textsubscript{x}}(O)    \textit{d\textsubscript{zz}}(Mn)    \textit{p\textsubscript{z}}(O)    \textit{d\textsubscript{xx}}(Mn)    
 & -0.561 & 2 & 0.200 & 0.168 & 0.322 & 0.251 \\
\hline
14 & \textit{p\textsubscript{x}}(O)    \textit{p\textsubscript{z}}(O)    \textit{p\textsubscript{y}}(O)    \textit{p\textsubscript{x}}(C)    
 & -0.554 & 2 & 0.115 & 0.138 & 0.134 & 0.090 \\
\hline
15 & \textit{p\textsubscript{y}}(C)    \textit{s}(H)    
 & -0.537 & 2 & 0.026 & 0.021 & 0.023 & 0.150 \\
\hline
16 & \textit{d\textsubscript{xz}}(Mn)    \textit{p\textsubscript{x}}(N)    
 & -0.533 & 2 & 0.076 & 0.159 & 0.240 & 0.374 \\
\hline
17 & \textit{p\textsubscript{z}}(N)    \textit{d\textsubscript{xy}}(Mn)    \textit{p\textsubscript{x}}(N)    \textit{p\textsubscript{x}}(C)    \textit{p\textsubscript{y}}(N)    \textit{p\textsubscript{z}}(O)    
 & -0.494 & 2 & 0.235 & 0.142 & 0.221 & 0.104 \\
\hline
18 & \textit{p\textsubscript{x}}(N)    \textit{p\textsubscript{z}}(N)    \textit{p\textsubscript{z}}(O)    \textit{p\textsubscript{x}}(O)    \textit{d\textsubscript{xx}}(Mn)    
 & -0.477 & 2 & 0.270 & 0.165 & 0.341 & 0.139 \\
\hline
19 & \textit{p\textsubscript{z}}(O)    \textit{s}(H)    \textit{p\textsubscript{z}}(C)    \textit{p\textsubscript{x}}(N)    
 & -0.474 & 2 & 0.113 & 0.023 & 0.029 & 0.132 \\
\hline
20 & \textit{p\textsubscript{x}}(O)    \textit{p\textsubscript{z}}(O)    \textit{d\textsubscript{yz}}(Mn)    \textit{s}(H)    
 & -0.471 & 2 & 0.160 & 0.020 & 0.026 & 0.037 \\
\hline
21 & \textit{p\textsubscript{z}}(O)    \textit{p\textsubscript{x}}(O)    \textit{s}(H)    \textit{d\textsubscript{zz}}(Mn)    
 & -0.442 & 2 & 0.073 & 0.020 & 0.028 & 0.038 \\
\hline
22 & \textit{p\textsubscript{x}}(Cl)    \textit{p\textsubscript{z}}(Cl)    \textit{d\textsubscript{xz}}(Mn)    \textit{s}(Mn)    
 & -0.390 & 2 & 0.122 & 0.274 & 0.698 & 0.237 \\
\hline
23 & \textit{p\textsubscript{y}}(Cl)    
 & -0.355 & 2 & 0.057 & 0.196 & 0.053 & 0.041 \\
\hline
24 & \textit{p\textsubscript{z}}(Cl)    \textit{p\textsubscript{x}}(Cl)    \textit{p\textsubscript{y}}(Cl)    
 & -0.350 & 2 & 0.036 & 0.081 & 0.005 & 0.010 \\
\hline
25 & \textit{p\textsubscript{x}}(C)    \textit{p\textsubscript{y}}(Cl)    \textit{p\textsubscript{z}}(C)    
 & -0.326 & 2 & 0.199 & 0.648 & 0.809 & 0.699 \\
\hline
26 & \textit{p\textsubscript{x}}(Cl)    \textit{p\textsubscript{z}}(Cl)    \textit{p\textsubscript{x}}(C)    \textit{p\textsubscript{z}}(C)    \textit{p\textsubscript{z}}(N)    
 & -0.322 & 2 & 0.167 & 0.666 & 0.805 & 0.523 \\
\hline
27 & \textit{d\textsubscript{zz}}(Mn)    \textit{d\textsubscript{xx}}(Mn)    \textit{p\textsubscript{x}}(O)    \textit{p\textsubscript{z}}(O)    \textit{d\textsubscript{xy}}(Mn)    
 & -0.070 & 0 & 0.807 & 0.312 & 0.204 & 0.280 \\
\hline
28 & \textit{d\textsubscript{xy}}(Mn)    \textit{d\textsubscript{yz}}(Mn)    \textit{d\textsubscript{xx}}(Mn)    \textit{p\textsubscript{y}}(O)    \textit{d\textsubscript{yy}}(Mn)    
 & 0.043 & 0 & 0.687 & 0.308 & 0.897 & 0.205 \\
\hline
29 & \textit{d\textsubscript{xz}}(Mn)    \textit{s}(Mn)    \textit{s}(C)    \textit{s}(H)    \textit{p\textsubscript{x}}(C)    \textit{p\textsubscript{z}}(C)    \textit{p\textsubscript{z}}(N)    
 & 0.088 & 0 & 0.223 & 0.267 & 0.088 & 0.210 \\
\hline
30 & \textit{s}(Mn)    \textit{d\textsubscript{xz}}(Mn)    \textit{p\textsubscript{z}}(C)    \textit{p\textsubscript{x}}(C)    \textit{p\textsubscript{z}}(N)    \textit{d\textsubscript{xx}}(Mn)    \textit{s}(N)    
 & 0.094 & 0 & 0.203 & 0.149 & 0.131 & 0.124 \\
\hline
31 & \textit{d\textsubscript{xz}}(Mn)    \textit{p\textsubscript{z}}(C)    \textit{s}(Mn)    \textit{p\textsubscript{x}}(N)    \textit{p\textsubscript{x}}(C)    
 & 0.099 & 0 & 0.218 & 0.167 & 0.189 & 0.140 \\
\hline
32 & \textit{s}(Mn)    \textit{s}(C)    \textit{d\textsubscript{xz}}(Mn)    \textit{s}(Cl)    
 & 0.128 & 0 & 0.054 & 0.043 & 0.027 & 0.054 \\
\hline
33 & \textit{d\textsubscript{yz}}(Mn)    \textit{d\textsubscript{xy}}(Mn)    \textit{s}(H)    \textit{d\textsubscript{xz}}(Mn)    \textit{d\textsubscript{yy}}(Mn)    \textit{s}(N)    
 & 0.139 & 0 & 0.225 & 0.802 & 0.591 & 0.581 \\
\hline
34 & \textit{s}(C)    \textit{s}(H)    \textit{s}(Mn)    
 & 0.179 & 0 & 0.014 & 0.031 & 0.032 & 0.038 \\
\hline
\end{tabular}}
\end{table}

\begin{figure}[ht]
\includegraphics[width=12cm]{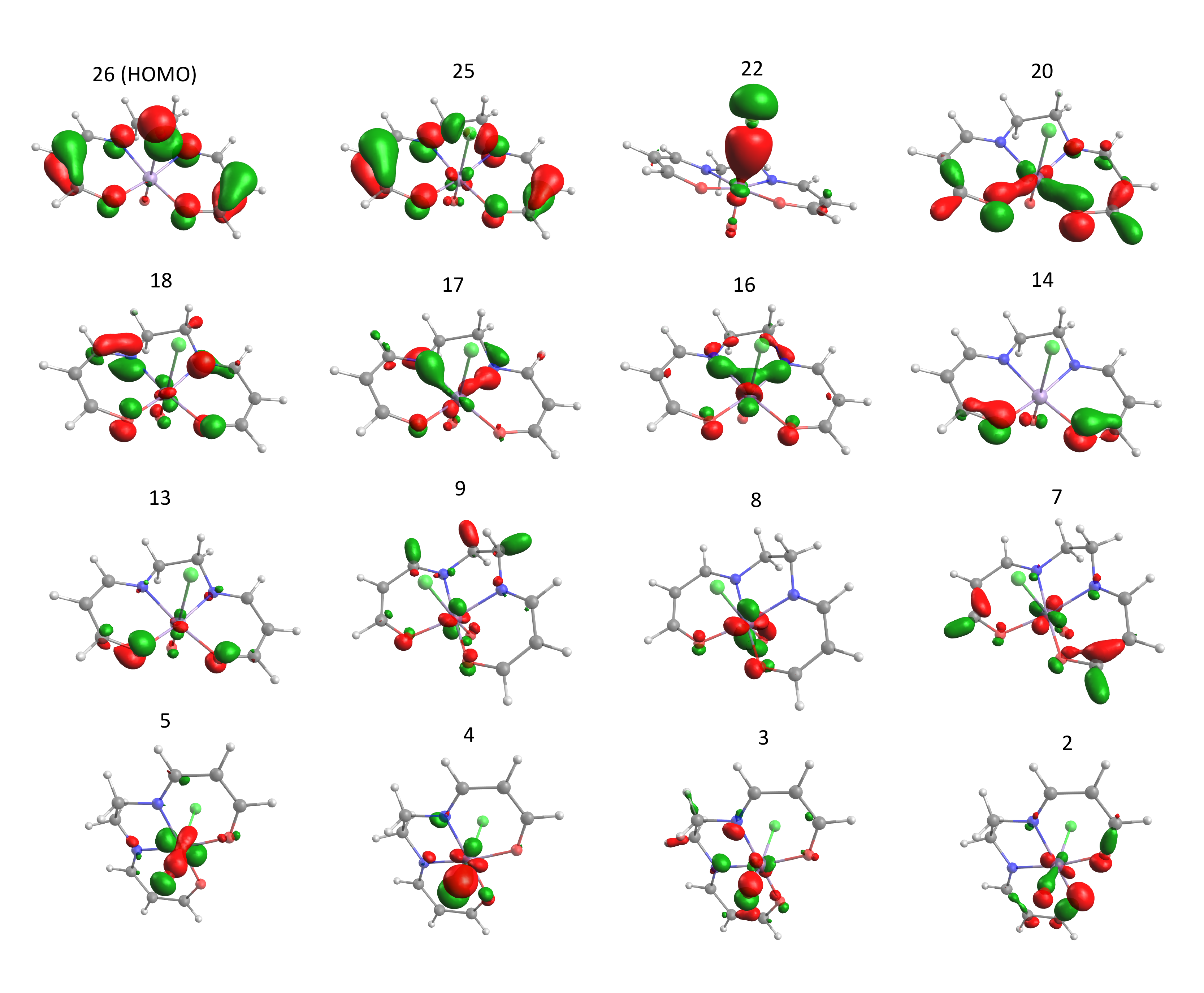}
\caption{The selected occupied molecular orbitals for oxo-Mn(salen). The numbering is in accordance with the numbering of orbitals in table \ref{table: oxosalen}.}
\label{fig:oxosalen}
\centering
\end{figure}
\begin{figure}[ht]
\includegraphics[width=12cm]{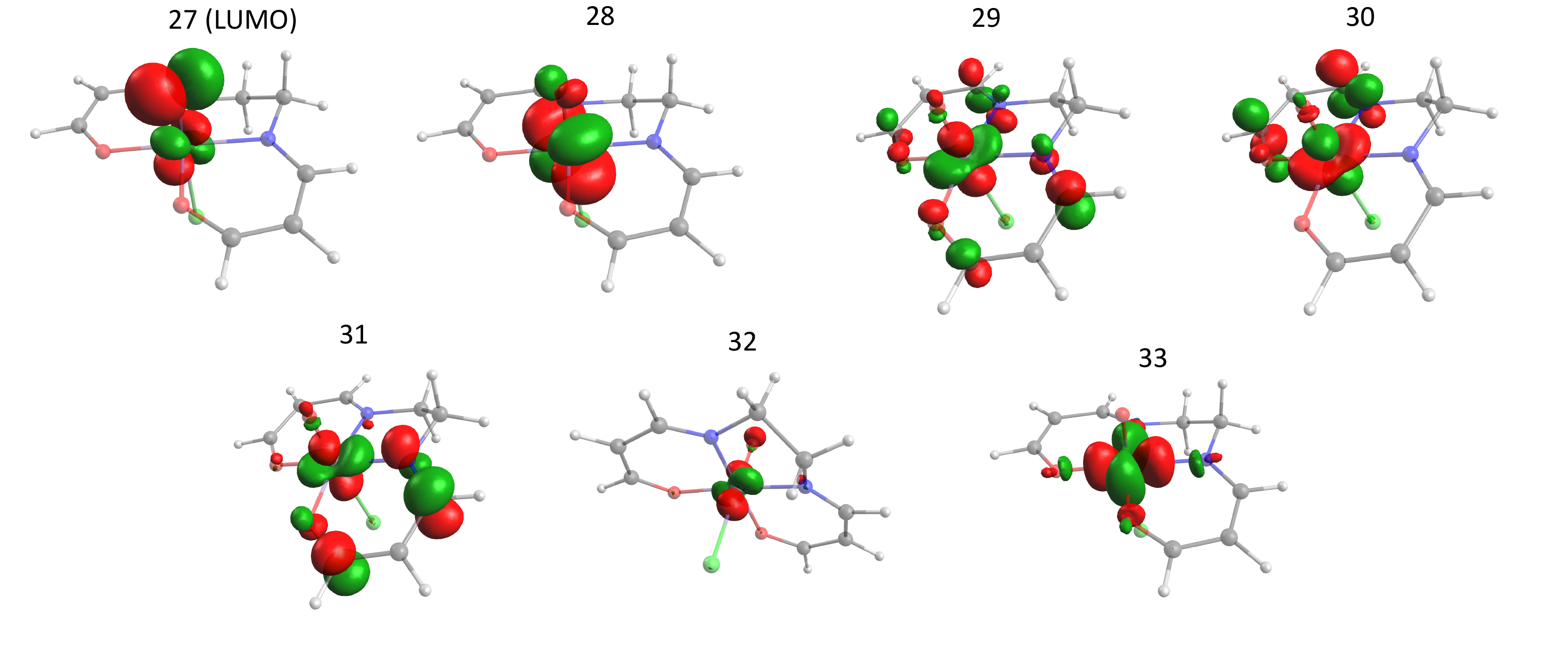}
\caption{The selected virtual molecular orbitals for oxo-Mn(salen). The numbering is in accordance with the numbering of orbitals in table \ref{table: oxosalen}.}
\label{fig:oxosalen_v}
\centering
\end{figure}

\textbf{Oxo-Mn(salen).} Oxo-Mn(salen) is another important organometallic molecule, able to catalyse the asymmetric epoxidation of olefins\cite{Zhang-Loebach-1990, Irie-Noda-1990}. The lowest singlet and triplet states are nearly degenerated, with the singlet one being a ground state in a margin of around 0.5 kcal/mol\cite{Sears-Sherrill-2006, Olivares-Amaya-Hu-2015, Stein-Reiher-2016, Antalik-Veis-2019}. Like for Fe(II)-porphyrin the choice of the active space is extremely important, since triplet state also might be predicted as the lowest one\cite{Ma-Manni-2011,Wouters-Bogaerts-2014}. Manual selection scheme of the content of the active space in minimal form involve picking non-bondign \textit{3d\textsubscript{x\textsuperscript{2}--y\textsuperscript{2}}} (Mn) orbital, bonding/antibonding orbitals with the mix of \textit{d} (Mn) and \textit{2p} (O/N) atomic orbitals (to account for Mn-O/N $\sigma$ bonding), $\sigma$ and $\pi$ bonding/antibonding orbitals of the O in axial position to Mn, $\pi$ bonding/antibonding orbitals from the salen ligand\cite{Ivanic-Collins-2004, Ma-Manni-2011,Wouters-Bogaerts-2014}. It can be extended by \textit{4d}, \textit{4p}, \textit{4s} orbitals on Mn and \textit{3p} orbitals on Cl\cite{Olivares-Amaya-Hu-2015, Antalik-Veis-2019}.

In our calculations we can distinguish 21 orbitals with single-entropies larger than 0.1, and 8 orbitals with single-entropies in the range from 0.05 to 0.1. This is consistent with results of the automated selection scheme by Stein and Reiher that selects the space of the size 26\cite{Abrams-Sherrill-2004}. Model 5x896 performs as 17/20, 18/22, 19/24, and 21/26. The most serious misinterpretation occurs for orbital 9 
-- combination of \textit{d} (Mn), \textit{p} (O/H), and \textit{s} (H) atomic orbitals. Seemingly, that happens because of the \textit{s} (H) atomic orbitals, which are rarely involved as building blocks for strongly correlated molecular orbitals and often appear with insignificant ones, thus NNs might have tendency to understated the corresponding weights. Contrary, the similar molecular orbital number 8, which is interpreted as a strongly correlated one by all models, does not have \textit{s} (H) contributions. The same problem with \textit{s} (H) contributions concerns the orbital 10 and 20. In the same time it should be noted, that the used bond dimension 2048 might not be enough to properly describe oxo-Mn(salen) and with higher bond dimensions these problematic orbitals might be interpreted as less significant. Model 8x896at performs as 15/20, 16/22, 19/24, and 22/26 -- that is completely comparable to the 5x896 model. Results for model 5x128 are 17/20, 18/22, 19/24, and 20/26.

\begin{table}[h!]
\centering
\caption{Single-orbital entropies for trans-[Co(C\textsubscript{2}H\textsubscript{8}N\textsubscript{2})\textsubscript{2}(NO\textsubscript{2})Cl]\textsuperscript{+} -- true (DMRG) and predicted by models (5x128, 5x896, 5x896at). The table is cut to show only the most important part of the complete space. The full table can be found in Supporting Materials (Table S13).}
\label{table: coen2no2cl}
\resizebox{\textwidth}{!}{
\begin{tabular}  {C{0.04\textwidth} C{0.44\textwidth} C{0.1\textwidth} C{0.06\textwidth} C{0.08\textwidth} C{0.08\textwidth} C{0.08\textwidth} C{0.08\textwidth}} 
\hline
\textbf{Orb} & \textbf{Atomic orbitals} & \textbf{Energy} & \textbf{Occ} & \textbf{DMRG} & \textbf{$\textrm{5x128}$} & \textbf{$\textrm{5x896}$} & \textbf{$\textrm{5x896at}$} \\
\hline
6 & \textit{d\textsubscript{xy}}(Co)    \textit{d\textsubscript{xz}}(Co)    \textit{d\textsubscript{xx}}(Co)    \textit{d\textsubscript{yz}}(Co)    \textit{p\textsubscript{x}}(N)    
 & -0.919 & 2 & 0.076 & 0.139 & 0.096 & 0.092 \\
\hline
7 & \textit{d\textsubscript{yz}}(Co)    \textit{d\textsubscript{xz}}(Co)    \textit{d\textsubscript{xy}}(Co)    
 & -0.914 & 2 & 0.046 & 0.064 & 0.038 & 0.083 \\
\hline
8 & \textit{d\textsubscript{xz}}(Co)    \textit{d\textsubscript{xy}}(Co)    \textit{d\textsubscript{zz}}(Co)    \textit{p\textsubscript{x}}(N)    
 & -0.913 & 2 & 0.041 & 0.177 & 0.199 & 0.080 \\
\hline
9 & \textit{p\textsubscript{x}}(N)    \textit{p\textsubscript{z}}(O)    \textit{s}(O)    
 & -0.900 & 2 & 0.024 & 0.026 & 0.026 & 0.015 \\
\hline
10 & \textit{d\textsubscript{yz}}(Co)    \textit{p\textsubscript{x}}(N)    \textit{s}(H)    
 & -0.868 & 2 & 0.033 & 0.020 & 0.023 & 0.055 \\
\hline
11 & \textit{d\textsubscript{xy}}(Co)    \textit{d\textsubscript{xz}}(Co)    \textit{p\textsubscript{z}}(N)    
 & -0.863 & 2 & 0.041 & 0.141 & 0.037 & 0.206 \\
\hline
12 & \textit{d\textsubscript{xz}}(Co)    \textit{d\textsubscript{xy}}(Co)    \textit{d\textsubscript{xx}}(Co)    \textit{d\textsubscript{yz}}(Co)    
 & -0.857 & 2 & 0.075 & 0.132 & 0.029 & 0.083 \\
\hline
13 & \textit{p\textsubscript{x}}(N)    \textit{s}(H)    
 & -0.852 & 2 & 0.031 & 0.023 & 0.024 & 0.040 \\
\hline
14 & \textit{d\textsubscript{xy}}(Co)    \textit{d\textsubscript{xz}}(Co)    \textit{p\textsubscript{x}}(N)    
 & -0.843 & 2 & 0.040 & 0.073 & 0.027 & 0.102 \\
\hline
15 & \textit{d\textsubscript{yz}}(Co)    
 & -0.837 & 2 & 0.053 & 0.161 & 0.024 & 0.085 \\
\hline
16 & \textit{d\textsubscript{xz}}(Co)    \textit{d\textsubscript{xy}}(Co)    \textit{s}(O)    \textit{p\textsubscript{z}}(N)    \textit{p\textsubscript{x}}(O)    \textit{p\textsubscript{y}}(N)    \textit{p\textsubscript{z}}(O)    \textit{p\textsubscript{y}}(O)    
 & -0.820 & 2 & 0.020 & 0.254 & 0.140 & 0.133 \\
\hline
17 & \textit{p\textsubscript{y}}(N)    \textit{p\textsubscript{z}}(N)    \textit{d\textsubscript{xy}}(Co)    \textit{p\textsubscript{y}}(O)    \textit{p\textsubscript{z}}(O)    \textit{d\textsubscript{xz}}(Co)    
 & -0.802 & 2 & 0.088 & 0.137 & 0.161 & 0.257 \\
\hline
18 & \textit{d\textsubscript{yz}}(Co)    \textit{p\textsubscript{z}}(C)    \textit{p\textsubscript{z}}(N)    
 & -0.770 & 2 & 0.045 & 0.115 & 0.039 & 0.190 \\
\hline
19 & \textit{p\textsubscript{y}}(N)    
 & -0.764 & 2 & 0.071 & 0.120 & 0.037 & 0.137 \\
\hline
20 & \textit{p\textsubscript{z}}(N)    \textit{p\textsubscript{x}}(N)    \textit{p\textsubscript{y}}(N)    \textit{d\textsubscript{xz}}(Co)    
 & -0.764 & 2 & 0.041 & 0.171 & 0.051 & 0.287 \\
\hline
21 & \textit{p\textsubscript{y}}(C)    \textit{p\textsubscript{x}}(N)    \textit{p\textsubscript{y}}(N)    \textit{p\textsubscript{x}}(C)    
 & -0.758 & 2 & 0.045 & 0.135 & 0.032 & 0.167 \\
\hline
22 & \textit{p\textsubscript{z}}(C)    
 & -0.751 & 2 & 0.035 & 0.139 & 0.029 & 0.147 \\
\hline
23 & \textit{p\textsubscript{y}}(C)    \textit{p\textsubscript{z}}(C)    \textit{p\textsubscript{z}}(N)    
 & -0.724 & 2 & 0.038 & 0.146 & 0.043 & 0.210 \\
\hline
24 & \textit{p\textsubscript{x}}(C)    \textit{s}(H)    
 & -0.673 & 2 & 0.032 & 0.026 & 0.028 & 0.164 \\
\hline
25 & \textit{p\textsubscript{x}}(C)    \textit{p\textsubscript{y}}(C)    \textit{s}(H)    \textit{p\textsubscript{z}}(C)    
 & -0.672 & 2 & 0.036 & 0.026 & 0.029 & 0.174 \\
\hline
26 & \textit{p\textsubscript{y}}(N)    \textit{p\textsubscript{z}}(N)    \textit{d\textsubscript{yy}}(Co)    \textit{d\textsubscript{zz}}(Co)    \textit{s}(H)    
 & -0.666 & 2 & 0.137 & 0.020 & 0.037 & 0.090 \\
\hline
27 & \textit{p\textsubscript{z}}(C)    \textit{s}(H)    \textit{p\textsubscript{y}}(C)    
 & -0.663 & 2 & 0.044 & 0.033 & 0.028 & 0.210 \\
\hline
28 & \textit{p\textsubscript{y}}(N)    \textit{p\textsubscript{z}}(N)    
 & -0.655 & 2 & 0.038 & 0.137 & 0.204 & 0.310 \\
\hline
29 & \textit{p\textsubscript{z}}(N)    \textit{p\textsubscript{y}}(N)    \textit{p\textsubscript{z}}(C)    \textit{p\textsubscript{y}}(C)    
 & -0.631 & 2 & 0.039 & 0.222 & 0.408 & 0.403 \\
\hline
30 & \textit{p\textsubscript{x}}(O)    \textit{p\textsubscript{z}}(O)    \textit{p\textsubscript{y}}(O)    
 & -0.606 & 2 & 0.024 & 0.140 & 0.273 & 0.067 \\
\hline
31 & \textit{p\textsubscript{x}}(Cl)    \textit{p\textsubscript{x}}(O)    \textit{d\textsubscript{xx}}(Co)    \textit{s}(N)    
 & -0.594 & 2 & 0.197 & 0.268 & 0.399 & 0.482 \\
\hline
32 & \textit{p\textsubscript{y}}(O)    \textit{p\textsubscript{z}}(O)    
 & -0.566 & 2 & 0.161 & 0.124 & 0.216 & 0.059 \\
\hline
33 & \textit{p\textsubscript{x}}(Cl)    \textit{p\textsubscript{x}}(O)    \textit{p\textsubscript{x}}(N)    \textit{s}(Co)    \textit{p\textsubscript{x}}(Co)    \textit{s}(N)    
 & -0.552 & 2 & 0.055 & 0.135 & 0.536 & 0.108 \\
\hline
34 & \textit{p\textsubscript{y}}(Cl)    \textit{p\textsubscript{z}}(Cl)    
 & -0.541 & 2 & 0.014 & 0.125 & 0.285 & 0.201 \\
\hline
35 & \textit{p\textsubscript{z}}(Cl)    \textit{p\textsubscript{y}}(Cl)    \textit{p\textsubscript{x}}(Cl)    
 & -0.539 & 2 & 0.015 & 0.130 & 0.295 & 0.230 \\
\hline
36 & \textit{s}(Co)    \textit{d\textsubscript{xx}}(Co)    \textit{d\textsubscript{xy}}(Co)    \textit{d\textsubscript{yy}}(Co)    \textit{s}(N)    \textit{d\textsubscript{xz}}(Co)    \textit{d\textsubscript{zz}}(Co)    \textit{s}(Cl)    
 & -0.025 & 0 & 0.213 & 0.355 & 0.403 & 0.582 \\
\hline
37 & \textit{p\textsubscript{y}}(N)    \textit{p\textsubscript{z}}(N)    \textit{s}(C)    \textit{p\textsubscript{y}}(O)    
 & -0.016 & 0 & 0.277 & 0.752 & 0.659 & 0.317 \\
\hline
38 & \textit{s}(Co)    \textit{s}(Cl)    \textit{d\textsubscript{xx}}(Co)    \textit{s}(N)    \textit{s}(C)    
 & -0.009 & 0 & 0.166 & 0.437 & 0.148 & 0.258 \\
\hline
39 & \textit{d\textsubscript{yy}}(Co)    \textit{d\textsubscript{zz}}(Co)    \textit{s}(C)    \textit{d\textsubscript{xy}}(Co)    \textit{s}(N)    
 & -0.005 & 0 & 0.214 & 0.628 & 0.319 & 0.537 \\
\hline
40 & \textit{s}(C)    \textit{s}(N)    \textit{s}(H)    
 & 0.036 & 0 & 0.030 & 0.039 & 0.037 & 0.156 \\
\hline
41 & \textit{s}(C)    \textit{s}(H)    \textit{s}(N)    
 & 0.039 & 0 & 0.028 & 0.031 & 0.036 & 0.153 \\
\hline
42 & \textit{s}(C)    \textit{s}(H)    \textit{p\textsubscript{z}}(C)    
 & 0.055 & 0 & 0.032 & 0.027 & 0.032 & 0.173 \\
\hline
43 & \textit{s}(C)    \textit{s}(H)    \textit{s}(N)    
 & 0.066 & 0 & 0.029 & 0.026 & 0.035 & 0.133 \\
\hline
44 & \textit{s}(C)    \textit{s}(H)    \textit{s}(N)    
 & 0.075 & 0 & 0.037 & 0.028 & 0.036 & 0.163 \\
\hline
45 & \textit{s}(H)    \textit{s}(C)    \textit{p\textsubscript{y}}(C)    
 & 0.081 & 0 & 0.053 & 0.025 & 0.032 & 0.140 \\
\hline
46 & \textit{s}(H)    \textit{s}(C)    \textit{s}(N)    
 & 0.101 & 0 & 0.034 & 0.029 & 0.037 & 0.168 \\
\hline
\end{tabular}}
\end{table}

\begin{figure}[ht]
\includegraphics[width=12cm]{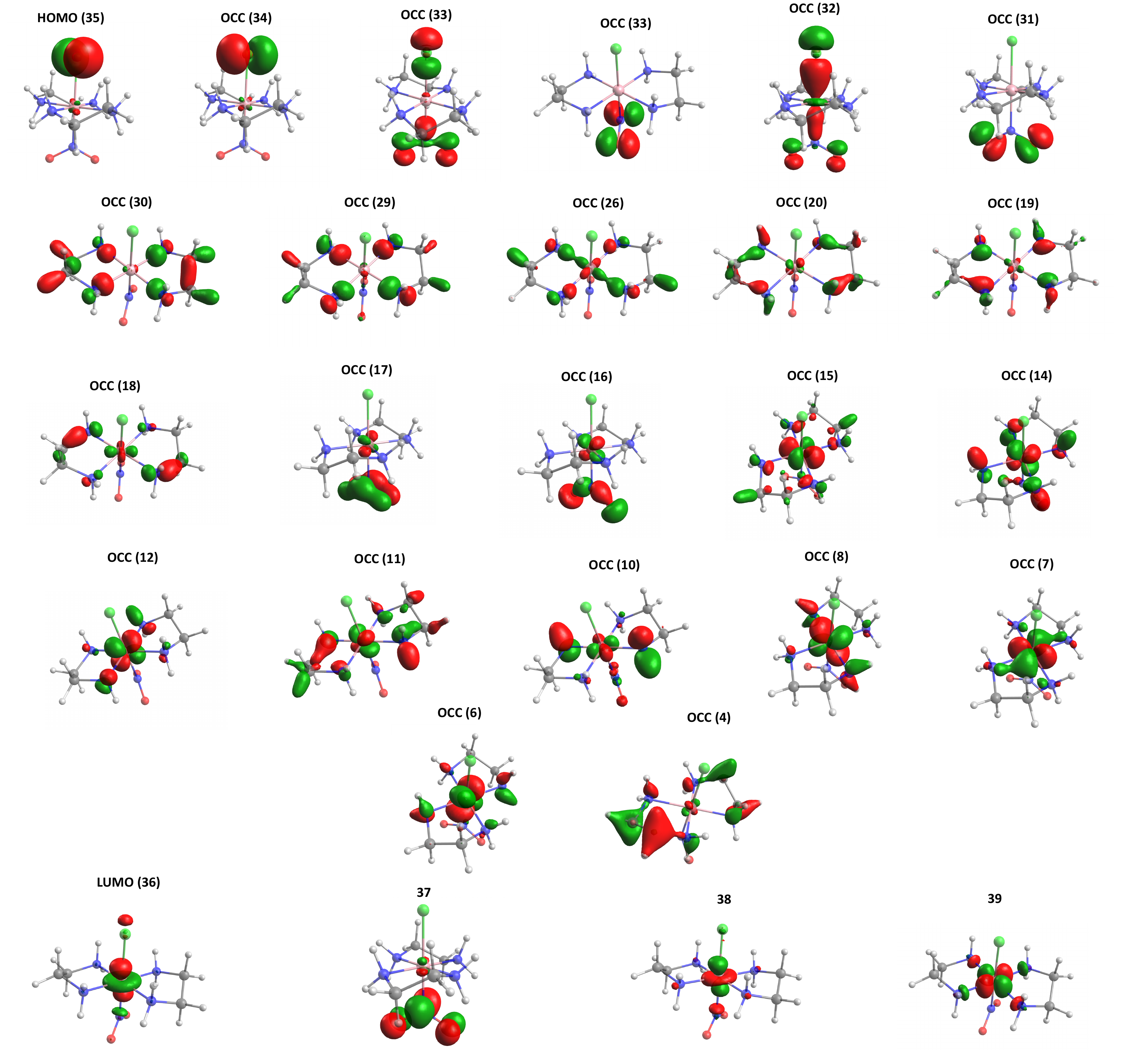}
\caption{The selected molecular orbitals for trans-[Co(C\textsubscript{2}H\textsubscript{8}N\textsubscript{2})\textsubscript{2}(NO\textsubscript{2})Cl]\textsuperscript{+}. The numbering is in accordance with the numbering of orbitals in table \ref{table: coen2no2cl}.}
\label{fig:coen2no2cl}
\centering
\end{figure}

\textbf{Trans-[Co(C\textsubscript{2}H\textsubscript{8}N\textsubscript{2})\textsubscript{2}(NO\textsubscript{2})Cl]\textsuperscript{+} (t-Co(En)\textsubscript{2}Cl(NO\textsubscript{2})).} This is CO(III) complex in octahedral coordination with two ethylenediamines, one Cl\textsuperscript{--} and one NO\textsubscript{2} groups as ligands. The most important orbitals are visualized in Figs.\ref{fig:coen2no2cl}. Our NN models tend to overestimate some orbitals with both \textit{d} (Co) and \textit{p} (NO\textsubscript{2}) characters (like orbitals 16 and 17, Table \ref{table: coen2no2cl}), or which are built from \textit{p} atomic orbitals on ethylenediamines (like orbitals 28 and 29, Table \ref{table: coen2no2cl}).
The predicted single-entropies for highly delocalised orbitals (for example, orbital 26, see Table \ref{table: coen2no2cl}) are underestimated with respect to DMRG. However, 6 most correlated orbitals (31, 32, 36 -- 39, Table \ref{table: coen2no2cl}) are identified correctly by 5x896, they all are included to the active space of the size 14. First 9 orbitals can be found in the active space of the size 22, which is apparently a reasonable size of the active space for t-Co(En)\textsubscript{2}Cl(NO\textsubscript{2}). Other NN models perform worse, since 5x128 and 5x896at seriously overestimate a lot of low lying occupied molecular orbitals (20 -- 23, Table \ref{table: coen2no2cl}), and, furthermore, 5x896at overestimates a chunk of virtual orbitals (40 -- 46, Table \ref{table: coen2no2cl}).

\newpage

\begin{table}[H]
\centering
\caption{Single-orbital entropies for [Fe\textsubscript{2}S\textsubscript{2}(SCH\textsubscript{3})]\textsuperscript{2--} -- true (DMRG) and predicted by models (5x128, 5x896, 5x896at). The table is cut to show only the most important part of the complete space. Label {\color{blue} B} marks bridging sulfur atoms, labels {\color{blue} 1} and {\color{blue} 2} mark different atoms in Fe\textsubscript{2}S\textsubscript{2} core. The full table can be found in Supporting Materials (Table S14).}
\label{table: fe2s2}
\resizebox{\textwidth}{!}{
\begin{tabular}  {C{0.04\textwidth} C{0.44\textwidth} C{0.1\textwidth} C{0.06\textwidth} C{0.08\textwidth} C{0.08\textwidth} C{0.08\textwidth} C{0.08\textwidth}} 
\hline
\textbf{Orb} & \textbf{Atomic orbitals} & \textbf{Energy} & \textbf{Occ} & \textbf{DMRG} & \textbf{$\textrm{5x128}$} & \textbf{$\textrm{5x896}$} & \textbf{$\textrm{5x896at}$} \\
\hline
1 & \textit{s}(S)    
 & -0.557 & 2 & 0.009 & 0.008 & 0.013 & 0.019 \\
\hline
2 & \textit{d\textsubscript{xx}}(Fe{\color{blue}\textsuperscript{1,2}})    \textit{d\textsubscript{yy}}(Fe{\color{blue}\textsuperscript{1,2}})    \textit{d\textsubscript{xz}}(Fe{\color{blue}\textsuperscript{1,2}})    
 & -0.374 & 2 & 0.062 & 0.339 & 0.409 & 0.494 \\
\hline
3 & \textit{d\textsubscript{yz}}(Fe{\color{blue}\textsuperscript{1,2}})    \textit{d\textsubscript{xy}}(Fe{\color{blue}\textsuperscript{1,2}})    
 & -0.363 & 2 & 0.044 & 0.443 & 0.506 & 0.503 \\
\hline
4 & \textit{d\textsubscript{yz}}(Fe{\color{blue}\textsuperscript{1,2}})    \textit{d\textsubscript{xy}}(Fe{\color{blue}\textsuperscript{1,2}})    
 & -0.356 & 2 & 0.042 & 0.505 & 0.585 & 0.621 \\
\hline
5 & \textit{d\textsubscript{xx}}(Fe{\color{blue}\textsuperscript{1,2}})    \textit{d\textsubscript{yy}}(Fe{\color{blue}\textsuperscript{1,2}})    \textit{d\textsubscript{xz}}(Fe{\color{blue}\textsuperscript{1,2}})    \textit{d\textsubscript{yz}}(Fe{\color{blue}\textsuperscript{1,2}})    \textit{s}(H)    
 & -0.349 & 2 & 0.044 & 0.027 & 0.053 & 0.177 \\
\hline
6 & \textit{s}(H)    \textit{p\textsubscript{y}}(C)    
 & -0.346 & 2 & 0.031 & 0.064 & 0.020 & 0.224 \\
\hline
7 & \textit{p\textsubscript{y}}(C)    \textit{d\textsubscript{yz}}(Fe{\color{blue}\textsuperscript{1,2}})    \textit{d\textsubscript{xx}}(Fe{\color{blue}\textsuperscript{1,2}})    \textit{s}(H)    \textit{d\textsubscript{xy}}(Fe{\color{blue}\textsuperscript{1,2}})    
 & -0.344 & 2 & 0.039 & 0.210 & 0.033 & 0.317 \\
\hline
8 & \textit{d\textsubscript{yz}}(Fe{\color{blue}\textsuperscript{1,2}})    \textit{s}(H)    \textit{p\textsubscript{x}}(C)    
 & -0.340 & 2 & 0.024 & 0.306 & 0.038 & 0.328 \\
\hline
9 & \textit{d\textsubscript{yz}}(Fe{\color{blue}\textsuperscript{1,2}})    \textit{p\textsubscript{x}}(C)    \textit{s}(H)    
 & -0.339 & 2 & 0.024 & 0.016 & 0.025 & 0.235 \\
\hline
10 & \textit{s}(H)    \textit{p\textsubscript{x}}(C)    \textit{p\textsubscript{y}}(C)    
 & -0.338 & 2 & 0.021 & 0.065 & 0.022 & 0.202 \\
\hline
11 & \textit{p\textsubscript{z}}(C)    \textit{d\textsubscript{yz}}(Fe{\color{blue}\textsuperscript{1,2}})    \textit{s}(H)    
 & -0.332 & 2 & 0.023 & 0.018 & 0.030 & 0.157 \\
\hline
12 & \textit{p\textsubscript{z}}(C)    \textit{d\textsubscript{yz}}(Fe{\color{blue}\textsuperscript{1,2}})    \textit{s}(H)    
 & -0.332 & 2 & 0.020 & 0.018 & 0.030 & 0.140 \\
\hline
13 & \textit{d\textsubscript{xx}}(Fe{\color{blue}\textsuperscript{1,2}})    \textit{d\textsubscript{yy}}(Fe{\color{blue}\textsuperscript{1,2}})    \textit{d\textsubscript{xz}}(Fe{\color{blue}\textsuperscript{1,2}})    \textit{d\textsubscript{xy}}(Fe{\color{blue}\textsuperscript{1,2}})    
 & -0.331 & 2 & 0.045 & 0.771 & 0.703 & 0.690 \\
\hline
14 & \textit{d\textsubscript{xz}}(Fe{\color{blue}\textsuperscript{1,2}})    \textit{d\textsubscript{zz}}(Fe{\color{blue}\textsuperscript{1,2}})    \textit{d\textsubscript{yy}}(Fe{\color{blue}\textsuperscript{1,2}})    \textit{d\textsubscript{xx}}(Fe{\color{blue}\textsuperscript{1,2}})    
 & -0.298 & 2 & 0.548 & 0.205 & 0.589 & 0.554 \\
\hline
15 & \textit{d\textsubscript{xz}}(Fe{\color{blue}\textsuperscript{1,2}})    \textit{d\textsubscript{zz}}(Fe{\color{blue}\textsuperscript{1,2}})    \textit{d\textsubscript{yy}}(Fe{\color{blue}\textsuperscript{1,2}})    
 & -0.282 & 2 & 0.489 & 0.352 & 0.640 & 0.358 \\
\end{tabular}}
\end{table}

\begin{table}[H]
\centering
\caption{Continuation: Single-orbital entropies for [Fe\textsubscript{2}S\textsubscript{2}(SCH\textsubscript{3})]\textsuperscript{2--} -- true (DMRG) and predicted by models (5x128, 5x896, 5x896at). The table is cut to show only the most important part of the complete space. Label {\color{blue} B} marks bridging sulfur atoms, labels {\color{blue} 1} and {\color{blue} 2} mark different atoms in Fe\textsubscript{2}S\textsubscript{2} core. The full table can be found in Supporting Materials (Table S14).}
\label{table: fe2s2_2}
\resizebox{\textwidth}{!}{
\begin{tabular}  {C{0.04\textwidth} C{0.44\textwidth} C{0.1\textwidth} C{0.06\textwidth} C{0.08\textwidth} C{0.08\textwidth} C{0.08\textwidth} C{0.08\textwidth}} 
\hline
\textbf{Orb} & \textbf{Atomic orbitals} & \textbf{Energy} & \textbf{Occ} & \textbf{DMRG} & \textbf{$\textrm{5x128}$} & \textbf{$\textrm{5x896}$} & \textbf{$\textrm{5x896at}$} \\
\hline
16 & \textit{p\textsubscript{z}}(C)    \textit{p\textsubscript{y}}(S)    \textit{p\textsubscript{y}}(C)    
 & -0.224 & 2 & 0.068 & 0.816 & 0.437 & 0.488 \\
\hline
17 & \textit{p\textsubscript{z}}(C)    \textit{p\textsubscript{z}}(S)    
 & -0.219 & 2 & 0.087 & 0.812 & 0.429 & 0.505 \\
\hline
18 & \textit{p\textsubscript{y}}(C)    \textit{p\textsubscript{y}}(S)    \textit{s}(S)    
 & -0.209 & 2 & 0.059 & 0.804 & 0.785 & 0.336 \\
\hline
19 & \textit{p\textsubscript{z}}(C)    \textit{p\textsubscript{y}}(C)    \textit{s}(S)    \textit{p\textsubscript{y}}(S)    
 & -0.207 & 2 & 0.059 & 0.450 & 0.825 & 0.384 \\
\hline
20 & \textit{p\textsubscript{x}}(S{\color{blue}\textsuperscript{B1,2}})
 & -0.178 & 2 & 0.247 & 0.740 & 0.686 & 0.421 \\
\hline
21 & \textit{p\textsubscript{z}}(S{\color{blue}\textsuperscript{B1,2}})  \textit{d\textsubscript{xx}}(Fe{\color{blue}\textsuperscript{1,2}})  
 & -0.170 & 2 & 0.175 & 0.773 & 0.682 & 1.010 \\
\hline
22 & \textit{p\textsubscript{y}}(S{\color{blue}\textsuperscript{B1,2}})  
 & -0.150 & 2 & 0.269 & 0.591 & 0.644 & 0.446 \\
\hline
23 & \textit{p\textsubscript{x}}(S{\color{blue}\textsuperscript{B1,2}}) 
 & -0.137 & 2 & 0.197 & 0.782 & 0.672 & 0.323 \\
\hline
24 & \textit{p\textsubscript{y}}(S)    \textit{p\textsubscript{x}}(S)  \textit{p\textsubscript{z}}(S{\color{blue}\textsuperscript{B1,2}})
 & -0.114 & 2 & 0.194 & 0.762 & 0.580 & 0.287 \\
\hline
25 & \textit{p\textsubscript{y}}(S{\color{blue}\textsuperscript{B1,2}}) \textit{p\textsubscript{x}}(S)    \textit{d\textsubscript{xy}}(Fe{\color{blue}\textsuperscript{1,2}})    
 & -0.106 & 2 & 0.198 & 0.799 & 0.792 & 0.797 \\
\hline
26 & \textit{p\textsubscript{y}}(S{\color{blue}\textsuperscript{B1,2}})  \textit{p\textsubscript{y}}(S)    \textit{p\textsubscript{x}}(S)    
 & -0.094 & 2 & 0.149 & 0.652 & 0.480 & 0.253 \\
\hline
27 & \textit{p\textsubscript{x}}(S)    \textit{p\textsubscript{y}}(S)    \textit{d\textsubscript{yz}}(Fe)    
 & -0.091 & 2 & 0.255 & 0.708 & 0.473 & 0.454 \\
\hline
28 & \textit{p\textsubscript{y}}(S{\color{blue}\textsuperscript{B1,2}}) \textit{p\textsubscript{y}}(S)    \textit{d\textsubscript{yz}}(Fe{\color{blue}\textsuperscript{1,2}})    
 & -0.089 & 2 & 0.113 & 0.670 & 0.551 & 0.530 \\
\hline
29 & \textit{p\textsubscript{z}}(S)    \textit{p\textsubscript{y}}(S)    
 & -0.069 & 2 & 0.026 & 0.703 & 0.851 & 0.195 \\
\hline
30 & \textit{p\textsubscript{x}}(S)    \textit{p\textsubscript{y}}(S)    \textit{p\textsubscript{z}}(S)    
 & -0.065 & 2 & 0.037 & 0.685 & 0.851 & 0.181 \\
\hline
31 & \textit{p\textsubscript{z}}(S)    \textit{p\textsubscript{x}}(S)    
 & -0.063 & 2 & 0.088 & 0.700 & 0.825 & 0.215 \\
\hline
32 & \textit{p\textsubscript{z}}(S)  \textit{p\textsubscript{x}}(S{\color{blue}\textsuperscript{B1,2}}) \textit{p\textsubscript{x}}(S)    \textit{d\textsubscript{xz}}(Fe{\color{blue}\textsuperscript{1,2}})    
 & -0.047 & 2 & 0.293 & 0.367 & 0.247 & 0.243 \\
\hline
33 & \textit{p\textsubscript{z}}(S{\color{blue}\textsuperscript{B1,2}})    \textit{d\textsubscript{xz}}(Fe{\color{blue}\textsuperscript{1,2}})    \textit{d\textsubscript{zz}}(Fe{\color{blue}\textsuperscript{1,2}})    \textit{s}(Fe{\color{blue}\textsuperscript{1,2}})    
 & 0.172 & 0 & 1.007 & 0.733 & 0.468 & 0.561 \\
\hline
34 & \textit{s}(Fe{\color{blue}\textsuperscript{1,2}})    \textit{s}(C)  \textit{s}(S)  \textit{s}(S{\color{blue}\textsuperscript{B1,2}}) 
 & 0.356 & 0 & 0.036 & 0.014 & 0.018 & 0.019 \\
\hline
35 & \textit{s}(Fe{\color{blue}\textsuperscript{1,2}})    \textit{s}(S)    \textit{s}(C)    
 & 0.370 & 0 & 0.021 & 0.011 & 0.017 & 0.014 \\
\hline
36 & \textit{s}(C)    \textit{s}(H)    \textit{d\textsubscript{xz}}(Fe{\color{blue}\textsuperscript{1,2}})    \textit{s}(Fe{\color{blue}\textsuperscript{1,2}})    
 & 0.377 & 0 & 0.272 & 0.122 & 0.082 & 0.194 \\
\hline
37 & \textit{s}(C)    \textit{d\textsubscript{xy}}(Fe{\color{blue}\textsuperscript{1,2}})    \textit{s}(H)    
 & 0.394 & 0 & 0.377 & 0.235 & 0.116 & 0.139 \\
\hline
38 & \textit{s}(C)    \textit{s}(H)    \textit{s}(Fe{\color{blue}\textsuperscript{1,2}})    
 & 0.402 & 0 & 0.084 & 0.015 & 0.028 & 0.032 \\
\hline
39 & \textit{s}(C)    \textit{s}(H)    \textit{s}(S)    
 & 0.412 & 0 & 0.077 & 0.012 & 0.019 & 0.028 \\
\hline
40 & \textit{s}(C)    \textit{s}(H)    
 & 0.413 & 0 & 0.110 & 0.015 & 0.025 & 0.037 \\
\hline
41 & \textit{s}(C)    \textit{s}(H)    \textit{s}(Fe{\color{blue}\textsuperscript{1,2}})    
 & 0.424 & 0 & 0.129 & 0.017 & 0.032 & 0.041 \\
\hline
42 & \textit{s}(C)    \textit{d\textsubscript{xy}}(Fe{\color{blue}\textsuperscript{1,2}})    \textit{s}(Fe{\color{blue}\textsuperscript{1,2}})    \textit{s}(H)    
 & 0.440 & 0 & 0.252 & 0.044 & 0.088 & 0.207 \\
\hline
43 & \textit{s}(H)    \textit{s}(C)    \textit{s}(Fe{\color{blue}\textsuperscript{1,2}})    \textit{d\textsubscript{xz}}(Fe{\color{blue}\textsuperscript{1,2}})    
 & 0.446 & 0 & 0.296 & 0.101 & 0.505 & 0.152 \\
\hline
44 & \textit{s}(H)    \textit{p\textsubscript{x}}(C)    
 & 0.469 & 0 & 0.057 & 0.015 & 0.025 & 0.033 \\
\hline
50 & \textit{s}(H)    \textit{p\textsubscript{z}}(C)    \textit{p\textsubscript{x}}(C)    
 & 0.502 & 0 & 0.063 & 0.019 & 0.023 & 0.031 \\
\hline
54 & \textit{s}(Fe)    \textit{p\textsubscript{z}}(C)    \textit{s}(H)    \textit{p\textsubscript{z}}(S)    
 & 0.540 & 0 & 0.055 & 0.024 & 0.027 & 0.031 \\
\hline
\end{tabular}}
\end{table}

\begin{figure}[h]
\includegraphics[width=12cm]{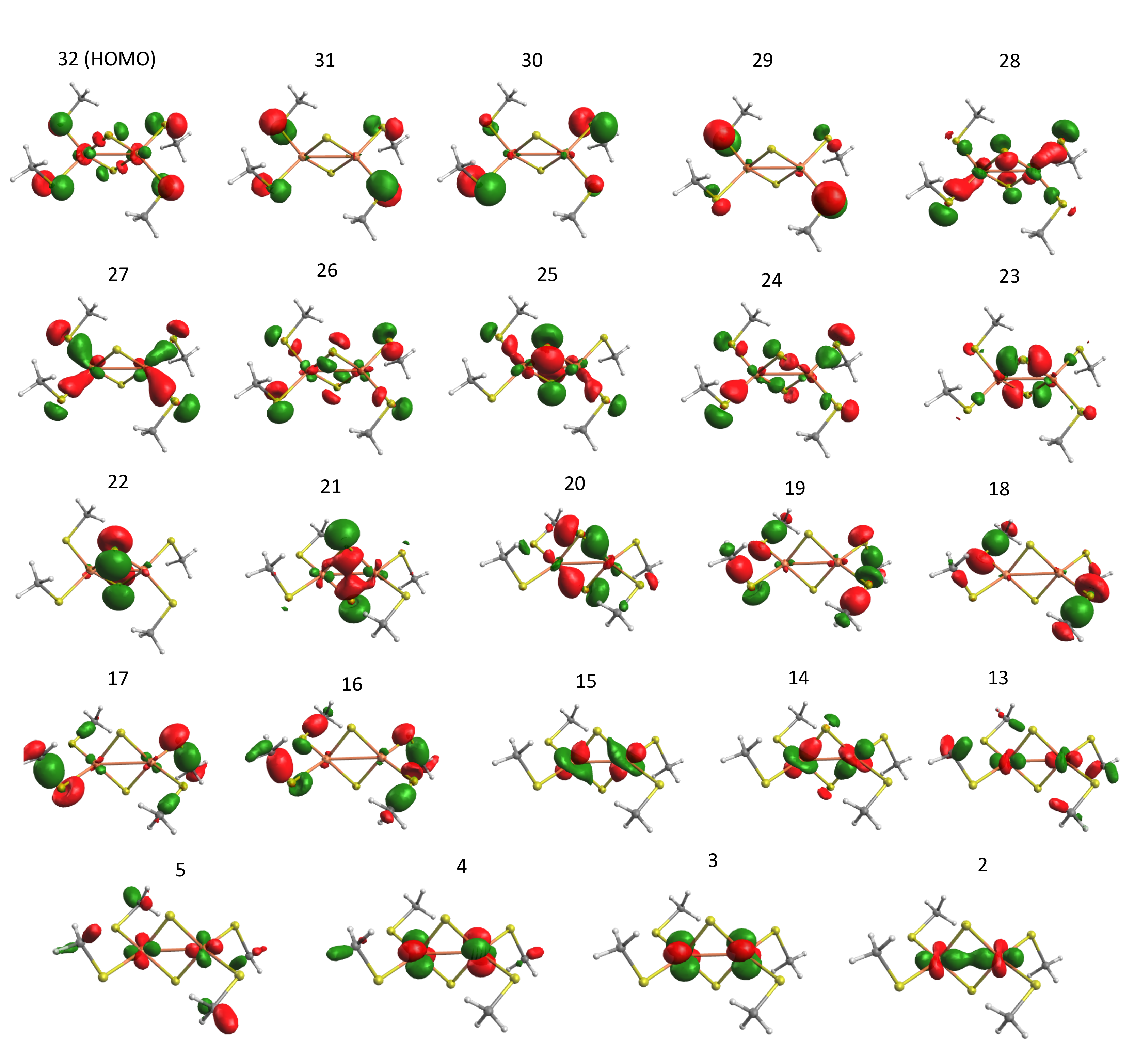}
\caption{The selected occupied molecular orbitals for [Fe$_2$S$_2$(SCH$_3$)]$^{2--}$ -- The numbering is in accordance with the numbering of orbitals in table \ref{table: fe2s2}}
\label{fig:fe2s2_o}
\centering
\end{figure}

\begin{figure}[h]
\includegraphics[width=12cm]{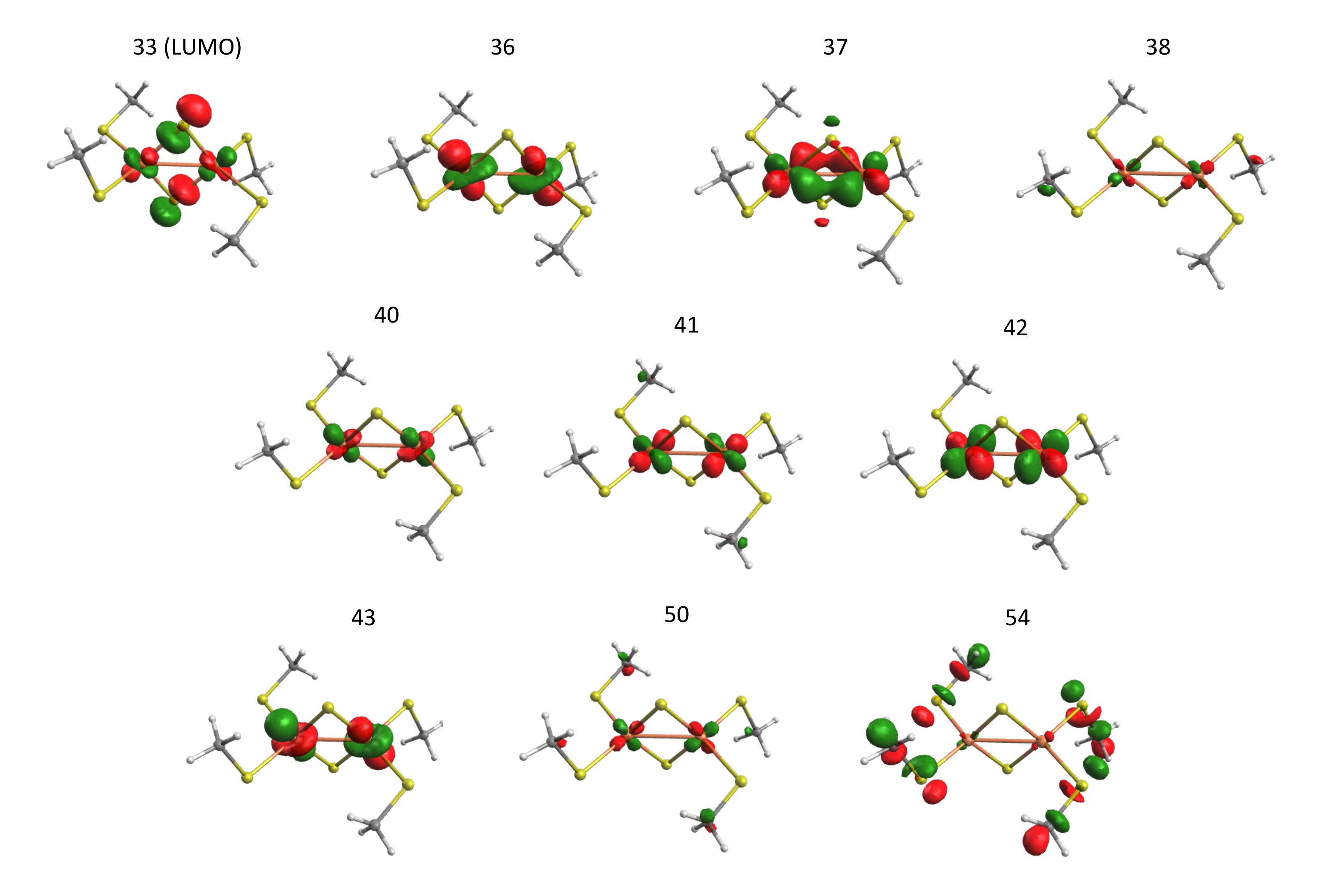}
\caption{The selected virtual molecular orbitals for [Fe\textsubscript{2}S\textsubscript{2}(SCH\textsubscript{3})]\textsuperscript{2--} -- The numbering is in accordance with the numbering of orbitals in table \ref{table: fe2s2}}
\label{fig:fe2s2_v}
\centering
\end{figure}

\textbf{[Fe\textsubscript{2}S\textsubscript{2}(SCH\textsubscript{3})\textsubscript{4}]\textsuperscript{2--}.} 
This is the most challenging testing system in this work.
The Fe\textsubscript{2}S\textsubscript{6} 2-- group is important part of many electron-transfer metalloproteins. The oxidation configuration --2 assumes Fe(III) spin states and its electronic structure is characterized by antiferromagnetic spin coupling of the metal centre, however the whole complex is singlet\cite{Sharma-Sivalingam-2014, Pandelia-Lanz-2015}. The proper description of the electronic structure requires a multiconfigurational approach since the static correlation is the main contributor to the correlation energy -- it affects not only the metal-ligand bonding but also describes the interaction of unpaired electrons on metal centres\cite{Sharma-Sivalingam-2014, Chu-Bovi-2017}. Depending on the applied active space solver, employed basis set and localization procedure, different active spaces have been chosen -- 10--16 (\textit{3d} orbitals of Fe/plus \textit{3p} orbitals of the bridging S) for complete CI calculations with single electron excitations out of several preselected sets of orbitals\cite{Presti-Stoneburner-2019, Kubas-2020}, and up to 32\cite{Sharma-Sivalingam-2014} or 36\cite{Cho-Rouxel-2019} with DMRG as active space solver.

The selected orbitals are visualized in Figs.\ref{fig:fe2s2_o} and \ref{fig:fe2s2_v}. From Tables \ref{table: fe2s2} and \ref{table: fe2s2_2} follow that in occupied space the low-lying $d-$orbitals 2-5 are overestimated, on the other hand even DMRG calculation show slightly elevated entropies as these $d$ components are important. The ML models correctly identifies $d$ occupied orbitals 14 and higher, and prefer also orbitals with mixed $p$ components from S and C, which are overestimated, but according to DMRG entropies still worth of consideration. The $p$- type orbitals centered on bridge sulphurs are predicted to be very important, aligning with an expectation as well as DMRG single-site entropies. The $p$ orbitals 29-31 of non-bridge sulphurs are overestimated, but for 31 the entropy of 0.088 from DMRG shows the importance of the orbital 31. The virtual space is predicted correctly for orbitals 33, 36-37, 42-43, but strongly delocalized orbitals with all weak components 38-41 are underestimated. 

In total in accordance to DMRG it is possible to distinguish 27 important and partially important orbitals (with $s^{(1)} > 0.05$). Model 5x896 catches 23 of them (85\%) when the active space of size 28 is selected, and 25 of them (approximately 92\%) when the active space of size 32 is selected.

\section{Conclusions}

We have developed NN models suited for very fast automatic selection of active spaces for strongly correlated systems. The models posses a nice degree of transferability, since they are able to predict majority of strongly correlated orbitals for molecules, which have not been included to the testing/training sets. In general, the model with 5 hidden layers, 896 neurons per hidden layer and \textbf{\textit{B}}+10tcEXC feature space shows the best performance, confirming the validity of the procedures for choosing of the models parameters and architecture. Feature space \textbf{\textit{B}}+10tcEXC has been chosen as abstract as possible, which increases the transferability level. We have shown that if models involve a feature space that brought the dependence on specific atom types (\textbf{\textit{B}}+10tcEXC+\textit{AT}), the performance is comparable or worse.

There is still a room for further improvement -- sometimes models experience problems with interpretation of the strongly delocalized orbitals without a clear character.
Also the procedure for optimization of the orbitals ordering is desirable, since the presented ordering from the orbital with the largest single-orbital entropy to the one with the smallest single-orbitals entropy is not optimal for DMRG calculations, and only serves for picking the most correlated orbitals to the active space of predefined size. The model that predicts two-orbital entropies would help in this case, however it requires different feature space and will be the subject of our follow up work.


\begin{acknowledgement}

This work has been supported by the Czech Science Foundation (grant no. 19-13126Y) and the Center for Scalable and Predictive methods for Excitation and Correlated phenomena (SPEC), which is funded by the U.S. Department of Energy (DOE), Office of Science, Office of Basic Energy Sciences, the Division of Chemical Sciences, Geosciences, and Biosciences. Also we would like to acknowledge the support by the Czech Ministry of Education, Youth and Sports from the Large Infrastructures for Research, Experimental Development and Innovations project “IT4Innovations National Supercomputing Center - LM2015070".

\end{acknowledgement}




\bibliography{Manuscript}

\end{document}